\documentclass[sigconf]{acmart}
\usepackage{booktabs} % For formal tables
\usepackage{natbib}
\usepackage{graphicx}
\usepackage{xspace}
\newcommand{\sysname}{\textit{DataProv}\xspace}

\acmDOI{10.475/123_4}

\acmISBN{123-4567-24-567/08/06}

\acmConference[CIKM'17]{ACM CIKM}{August 2017}{Singapore} 
\acmYear{2017}
\copyrightyear{2016}

\acmPrice{15.00}

\begin{document}
\settopmatter{printacmref=false} % Removes citation information below abstract
\renewcommand\footnotetextcopyrightpermission[1]{} % removes footnote with conference information in first column
\pagestyle{plain} % removes running headers
%MK:05-22-17:  Let us think about the name.
\title{Using Blockchain and smart contracts for secure data provenance management}
\author{Aravind Ramachandran}

\affiliation{%
  \institution{The University of Texas At Dallas}
  \streetaddress{800 W Campbell Rd}
  \city{Richardson} 
  \state{Texas} 
  \postcode{75080}
}
\email{axr156530@utdallas.edu}
\author{Dr.Murat Kantarcioglu}
\affiliation{%
  \institution{The University of Texas At Dallas}
  \streetaddress{800 W Campbell Rd}
  \city{Richardson} 
  \state{Texas} 
  \postcode{75080}
}
\email{ muratk@utdallas.edu}
\begin{abstract}
Blockchain technology has evolved from being an
immutable ledger of transactions for cryptocurrencies to a
programmable interactive environment for building distributed reliable
applications. Although, blockchain technology has been used
to address various challenges, to our knowledge none of the
previous work focused on using blockchain to develop a
secure and immutable scientific data provenance management framework that automatically  verifies the provenance records. In
this work, we leverage blockchain as a platform to facilitate
trustworthy data provenance collection, verification and management. The
developed system utilizes smart contracts and open provenance
model (OPM) to record immutable data trails. 
We show that our proposed framework can
efficiently and securely capture  and validate provenance data,  and  prevent any  malicious modification to  the captured data as long as  majority  of the participants are honest.
\end{abstract}

%
% The code below should be generated by the tool at
% http://dl.acm.org/ccs.cfm
% Please copy and paste the code instead of the example below. 
%

\keywords{Distributed systems,knowledge management, Data provenance, Block chain platform}
\maketitle

\section{Introduction}

As the data used for scientific research increases exponentially, ensuring information quality and preventing data manipulation has emerged as an important factor in validating the research results. For example, an audit conducted by the Cancer and Leukemia Group B, one of the multi-center cancer clinical trial groups sponsored by the National Cancer Institute, found an incidence of fraud of 0.25 percentage of the trials conducted~\cite{paper23}. 

To avoid data frauds such as data fabrication, under-reporting of the results and falsifying the results to match research objectives in critical scientific research, the provenance of the data has to be maintained. In this context, data provenance is defined as meta-data that describes where the data of interest originated, who owns the data and what were the transformations that were done to the data. Data provenance facilitates the integration of scientific data from diverse sources as well as providing verifiability of the sources. Also, it acts as a  yardstick for measuring how far the results of the experiments supports the actual objectives of the research and increases transparency and trustworthiness. For example, in~\cite{paper5}, authors highlight the increase in transparency and trustworthiness of research results due to data provenance tracking. Therefore, to increase transparency and trustworthiness, provenance details of the data must be recorded from its generation to the transformations to the productions of results. \footnote{In section 6, we discuss two real-world settings where the provenance of data is crucial to prevent fraud.} 

Main challenges for a provenance system are the collection and immutable storage of provenance data, verifiability and preserving the privacy of the collected provenance data. Although tracking data provenance is important, but equally important is to ensure that security and privacy of the collected provenance data is maintained. Data used in any form of research may come from a myriad of sources and may contain sensitive information such as patient records. Any form of data provenance management system should ensure that the data is protected against unauthorized access. Also, a data provenance system should guarantee that the provenance details recorded in it are verifiable by the authorized personal without compromising the privacy and violating the ownership of the data. 

Due this importance of collecting provenance information, systems such as Chimera\cite{paper29} and myGrid\cite{paper30} have been developed to store and process provenance information.  Many of the existing provenance systems are based on a centralized storage model. The downside to the centralized system architecture is that if the central server is compromised, the whole data provenance trails could be compromised. In provenance systems based on distributed architecture, the security of the data provenance information is another area of contention. Any authorized users can corrupt the data stored in the provenance system. To our knowledge, the current provenance systems do not try to validate the changes before they are stored. Our proposed \sysname  addresses these issues by using blockchain as a medium for storing provenance information and providing validations for each of the changes before logging the changes using smart contracts. The immutable nature of the blockchain environment ensure that the approved provenance changes cannot be modified by any users once they are stored. In  \sysname, due to the distributed nature of the blockchain, the data provenance trails are replicated on every node of the blockchain ensuring high availability and fault tolerance.

\subsection{Overview of Our Contributions} 
To address the above-mentioned challenges and requirements, in this work, we propose a system, \sysname, to securely capture scientific provenance data. \sysname combines the distributed immutable nature of the blockchain technology with cryptographic techniques to securely track data provenance without leaking privacy sensitive information. 
Furthermore, \sysname facilitates seamless generation of data provenance by authorized users and provides an automated method for verification of the generated provenance data. It also ensures the privacy of the data using public key encryption. The access control policies of the system restrict the access for the provenance data to authorized users.  

The \sysname eliminates the need for a trusted third party storage and verification of the provenance data using smart contracts and randomized voting process. Furthermore, monetary punishment mechanism is enabled to discourage any malicious changes. These monetary payment penalties are guaranteed to be enforced as long as half of the participants are honest.  Storage of the provenance data in \sysname is done using the log events of Smart Contracts~\cite{paper8} thereby saving further cost on storage. The \sysname system also provides customized verification scripts for authorized users to determine whether the changes submitted are valid or not.

We have implemented a \sysname system on top of Ethereum Blockchain~\cite{paper8} platform along with Meteor framework~\cite{paper22} for developing interfaces for the user's client module. The system was then evaluated in real world scenarios of clinical Drug trials and wheat production tracking system. The results show that \sysname system captures data provenance with fixed cost and moderate overheads.

The paper is structured as follows: Section\ref{sec:Background} describes the System model. Section \ref{sec:overview} discuses the system architecture and walks through the provenance life cycle. In Section\ref{sysdetil} we take a detailed look at the various components of the system and their functionalities. Section\ref{sec:vote} describes the two different types of the voting process implemented for verification of changes trails. In section\ref{sec:sysanalysis} , we analyze the security and privacy parameters of the system. Section\ref{sec:eval} details the results obtained by implementing \sysname in two real-world environments. In section \ref{sec:relatedwork} we compare \sysname with other related blockchain based systems. Section\ref{sec:conclude} we discuss the conclusions.
\section{Background}\label{sec:Background}
In this section, we discuss some of the  tools used by our system and our threat model assumptions.

\subsection{Ethereum}
\sysname is built on top of the  Ethereum, a distributed public blockchain network. Ethereum is a worldwide network of interconnected computers that execute and validate programs.  Ethereum provides a  decentralized Turing-complete platform called Ethereum virtual machines to run application codes called smart contracts. Ethereum also provides a currency called ether that is used to implement value exchange between nodes in the platform. Smart contracts are codes that reside within the Ethereum blockchain environment that executes when specific conditions are met. As the smart contracts reside on top of the ethereum blockchain, executions of the smart contract are also recorded in the blockchain. Smart contract can store and control ether. The functionality to control ether can be used to build applications that require deposit and payout of ethers such as online casino games,Identity managemets systems . In the Ethereum blockchain platform, each computational step has a cost associated with it~\cite{paper21} called gas.

\subsection{Provenance Model}
The \sysname system represents the data provenance trails using Open Provenance Model(OPM)~\cite{paper6}. In the OPM methodology, each action of the current system is represented using three parameters: 1) artifact (e.g., documents, files etc.) before and after change versions, 2) an agent which represents the initiator of the change and, 3) the process which is the process that changes the artifact from the previous version to the current version. 
 
In our project, we represent the OPM model as a triple describing what the agents, artifacts, and process are and also number coded relationship edges between them. For example, the action of modifying a file can be represented in OPM as a tuple (user, file: old version, file: new version, process used for modifications).

\subsection{Threat Model} \label{sec:threat} 
The \sysname system can have two types of attackers: an external adversary and an internal adversary. An external adversary is a user who does not have access to the document/data in the system, but will actively try to corrupt the data provenance trails of a particular private document/data. The external adversary does not know the key to decrypt the document nor does he have access to the location in which the document is stored. The adversary only has knowledge of the document id and uses this information to mount an attack on the blockchain based data provenance trail system. We assume that the cloud storage is not trustworthy. To overcome this vulnerability, we store the files in encrypted form.

An internal adversary has access to the document/data granted by the owner in the \sysname system. The internal adversary is able to change the document and log the changes as provenance trails on the blockchain. An internal advisory cannot grant access to a document to another (we assume the adversary is not the owner of the particular document). The internal adversary may use the access rights to corrupt the provenance trails by logging in  incorrect changes to the document trail. 
We assume that  at least half of the users that can access the  documents and associated provenance data are honest, and they can be trusted to verify  the correctness of the  changes done to the data.  We  believe that this  assumption is reasonable  since if most of the users are  malicious, we cannot  provide security guarantees~\cite{paper32}.

\section{System Overview}\label{sec:overview}
We consider a scientific research setting where researchers keep their research records as a document stored in the cloud. The document (e.g., any data file) is encrypted by the owner of the document (e.g., the lead researcher). Access to the document is restricted using public key encryption. The owner of the research document provides access to the document to users by providing the key. For a user to log the provenance information in the \sysname system, the owner of a document needs grant access to the document to the user. In the \sysname system model, the changes to the documents are made through versioning. Each change related to a document is stored as a separate new version. The system assumes that only the latest version of the document/data file is used for modification. The system checks the condition that any document which contains changes not logged in the provenance data is ignored. 

The system encourages truthful behavior by penalizing the users who submit wrong change provenance details. The voters are rewarded in the event they find a defective change submitted with a portion of the deposit amount for the change. The users log valid changes to the system using client applications running in each of the individual user's browser. Each of the client applications stores persistent data about the documents that the current user has access to using a back end database. For the current version of \sysname, meteor JS~\cite{paper21} and  MongoDB~\cite{paper25} are used to implement the client applications. The client applications communicate with the smart contract through a Geth node running at the client side. The smart contract system which stores the change records of the document is monitored for change events by the client. The smart contract stores access control policies along with details like the time of the last change of the particular document, signature of the last change and change logs etc.

\subsection{Provenance Capture Life Cycle}
\begin{figure}
  \includegraphics[width=8cm,height=6cm,scale=1]{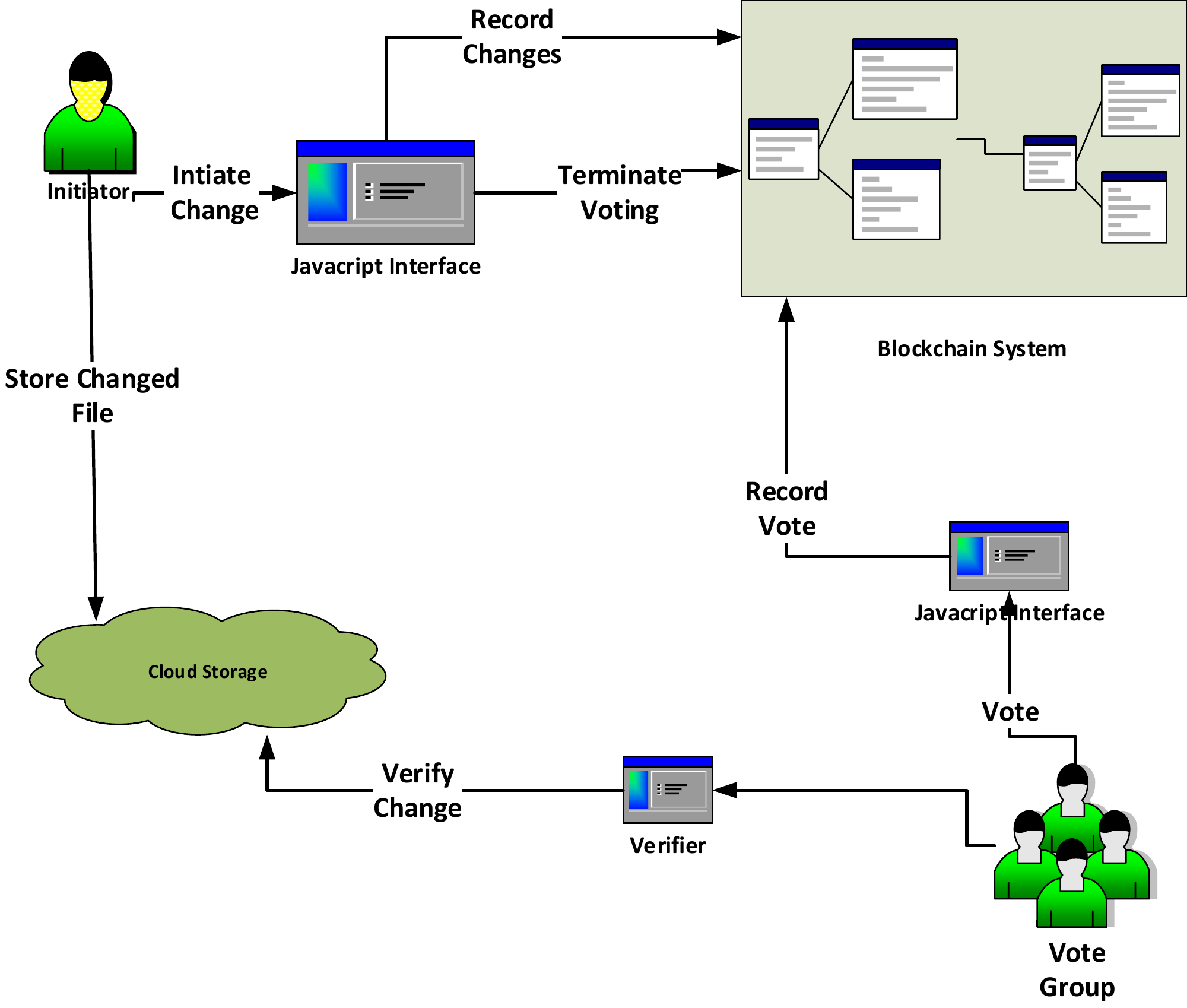}
  \caption{Provenance Life Cycle in \sysname System}
  \label{fig:life-cycle}
\end{figure}

An individual execution cycle of the \sysname system is shown in Figure \ref{fig:life-cycle}. The steps involve: 1) A user, who wishes to add or change the result set, modifies the latest version of the data file and then uploads it to the cloud server. Different versions of the documents/data files are maintained in the cloud so as to revert back in the event a change is rejected. 2) The change requester then submits a change request to Vote Contract through the client module along with a \textit{deposit amount}. The change request consists a digest comprising of: document Id, Encrypted form of  hashes of the previous and current versions of the data file, Link to the location of the file in the cloud repository, timestamp at which the change was made and also the signature of the requester. 3) The client module submits the change and then initiates the voting period. During the voting period, authorized user clients 
%changed to add the automated voting
verify the changes using the verifier script residing in the cloud storage. The scripts return true if the change is valid and false otherwise. 4) The clients cast their votes for/against the change based on the verification result, using the vote contract. The process is automated. 5) The vote contract records each of the votes cast by the users. At the end of the voting period, if the requisite amount of the users voted against the change, the change is rejected. The change initiator is penalized by the deposit amount and it is distributed among the voters. If after the voting period, the number of votes against the change is less than half of the voter, the change is accepted and the change requester is refunded the deposit amount. 6) In the event that a change is accepted after the voting process, the vote contract records the change in the document tracker contract. The log entries for each change consists of the following: author responsible for the change, the hash of the current document and the hash of the previous version of the document, high-level OPM representation of the current change and digital signature for future verification.

\section{System Details}\label{sysdetil}
The basic setup of \sysname system consists of two components. The on-chain components which mainly consist of Ethereum Smart contracts for access control, generating and storing provenance trails and conducting voting process, and the off-chain modules which consist of 
client application module that interfaces with the smart contract to submit the changes and keeps timers for the voting process and the cloud based script for verification of each of the data file changes that are submitted.

\subsection{On-chain module}
 The Ethereum blockchain platform provides executable programs that reside within the blockchain called Smart Contracts. The Smart contracts execute only when called and is capable of maintaining state variables.\sysname on-chain module mainly consists of two smart contracts  which  we discuss in details below.

 \subsubsection{Document Tracker contract}
 The Document Tracker smart contract is used to keep track of all the changes to a given document. Document Tracker contract implements access control policies and maintains all the user access information to the documents. The contract also provides methods for provenance trail generation for a  particular document. 
 The generated document trails are stored as events in the event log of Document Track contract. Event log storage of data provenance trails is preferred due to cost per storage consideration in Ethereum blockchain environment~\cite{paper26}.
 The format of the change event is described in detail in Appendix A. Each change event also stores the digital signature of the initiator based on the message digest. The document tracker supports all the basic functionality of access control management such as create a document for tracking, grant users rights to add changes to a particular document provenance history, revoke users access rights to a particular document history and finally generating and storing provenance history of a particular document to the log.  It is \textit{important to  note that }, \sysname \textit{does not  store any  sensitive information in plain text on the blockchain}, because any information stored on the blockchain including the smart contract code is publicly accessible. In addition, due to storage costs and blockchain storage limits, actual data is stored off the blockchain, potentially in a cloud location.  

The initial iteration of the provenance history is generated by the owner of a document when that document is added to the system. The contract enforces the constraint that granting access for adding provenance trails for a document is strictly controlled by the owner of the document. In the current implementation of \sysname, access rights to a particular document are nontransferable. In addition to the main methods, Document Tracker also consists of helper methods for checking the user access to a document and methods to update the owner of the document.  The Document track contract also implements checks to prevent unauthorized calls to the functions. Every provenance change event \textit{has to be approved through a voting process by the vote contract}. Access to the ChangeDocument method (e.g., log the change to a document) is therefore restricted to only a call from vote contract.  

We chose the  \textit{voting for verifying the submitted provenance information for two reasons}: 1) We want to efficiently prevent malicious changes that obviously violate data use constraints (e.g., not allowing the deletion of a patient record from the drug trial data). 2) We do not want the verification process to leak any sensitive information.  Unfortunately, verification process could  be very different in different settings. For example, for drug trials, main verification  process could be to make sure that no patient is deleted (e.g., to boost the success rate of the drug) from the data set due to a fatal reaction. Also, if the verification is done in the contract, we need to do this in a way that discloses no information (e.g., using zero knowledge proofs \cite{paper27} since contract source code and execution are publicly observable). To our knowledge, the existing zero knowledge techniques that are efficient are not general enough for all verification scenarios needed for our use case. At the same time, general zero knowledge verification techniques are not efficient enough to implement for provenance capturing \cite{paper28}. Due to these reasons, we allow each participant client progams to run the verification code off-the-chain and use on-the-chain contract to vote for or against the change. Below, we discuss the details of the voting process. 
y
\subsubsection{Vote contract} 
The vote contract implements the voting protocol. The contract implements two types of voting: simple majority voting and threshold voting which are discussed in section~\ref{sec:vote}. The initiator submits the change in an encrypted form along with his signature and document id to the vote contract. The vote contract receives the change and after verification generates a log event to initiate the voting phase for the change. The voting phase time interval for the set as $t_1$ ($t_1$ is set to one hour in our experiments) during which the participant can vote for/against the change. For each vote that is submitted, the vote contract verifies whether the vote is valid for the current voting period. At the end of the voting period, based on the type of voting process, the vote contract rejects/accepts the change based on the minimum number of  votes for or against the change. If the total votes for the current voting period do not reach the minimum threshold of the number votes required, the vote contract restarts the voting phase. At the end of the voting phase, if the decision is to accept the change, Vote contract submits the change to the Document Track contract for generating the provenance event. The vote contract currently accepts only a single outstanding change for a particular document for ensuring the continuity of the data provenance chain and consistency. The protocol also contains the option of logging changes without voting process for documents whose total user-base is less than three. 

\subsection{Off-chain module} 
The off chain client JavaScript module runs on the browser of each of the client machine. The JavaScript module acts as an interface between the user and back-end smart contracts. The client module is responsible for communicating with the smart contract for the storage of the changes, retrieval of the changes and verifying the validity of the changes. In addition to the client modules in each of the clients, \sysname also has a verification script module running at the cloud storage location where the different versions of the documents reside. The verification module verifies the validity of each change request of a particular document. 

The client modules consist of different components as discussed below:

\subsubsection{Client Interface module} The interface module mainly provides an interface for the user to interact with the smart contracts. The client module  provides access methods for all the basic operations of the system such as adding a new document for tracking, providing grant and revoke information and also tracking change trails for the documents. The Interface module implicitly generates the digital signature for all the operations that the user performs through the module.  
\subsubsection{Event Watcher module} The event watcher module observes the change events generated by the  Vote contract. The event watcher module reads any new change event logged into the contract, and checks if the current client is tracking the document for changes. If the current document change is relevant to the current user, the watcher contract decrypts the change event, verifies the signature of the change 
%changed to include the automated user voting
and then initiates a call to the verification script. It notifies the client about the verification results and if the verification result is valid, casts vote on behalf of the client. The voting process is automated such that the user need not be in the terminal. The watcher module uses a database to keep track of the details of documents that the current user is a stakeholder. 

\subsubsection{Timer module} The timer module is responsible for keeping track of the voting phases. When a change event to the document is generated, the timer module is called for initiating the voting interval for the change. The timer will trigger the termination of the voting process at the end of the voting interval. 

\subsubsection{Verification script:} The verification script resides within the cloud storage of the system. The verification script validates the data file/document changes that are submitted to the \sysname system. The input to the verification script includes current and previous cryptographic hash of the document (from the changes submitted to \sysname) and the link to the latest version of the file. The verification script first verifies whether the hashes submitted to the files are valid. It then compares the current unconfirmed data file with the last stable version of the file. If any other changes to the file other than the ones mentioned in the change request is identified, the verification script notifies the user of a mismatch. If there are no invalid changes in the file, the verification script confirms the change as valid to the user. Once the changes are verified as valid, to prevent further manipulation of the document version, the verification script restricts the write access to only the owner of the original document. The verification script \textit{can be customized according to the usage scenario} of the \sysname system and is developed as a plug-in module. In current implementation, we have developed the verification script based on Google appscript~\cite{paper24} to support Google Drive Storage.

\section{Voting Process} \label{sec:vote}
\begin{figure}
  \includegraphics[width=8cm,height=6cm]{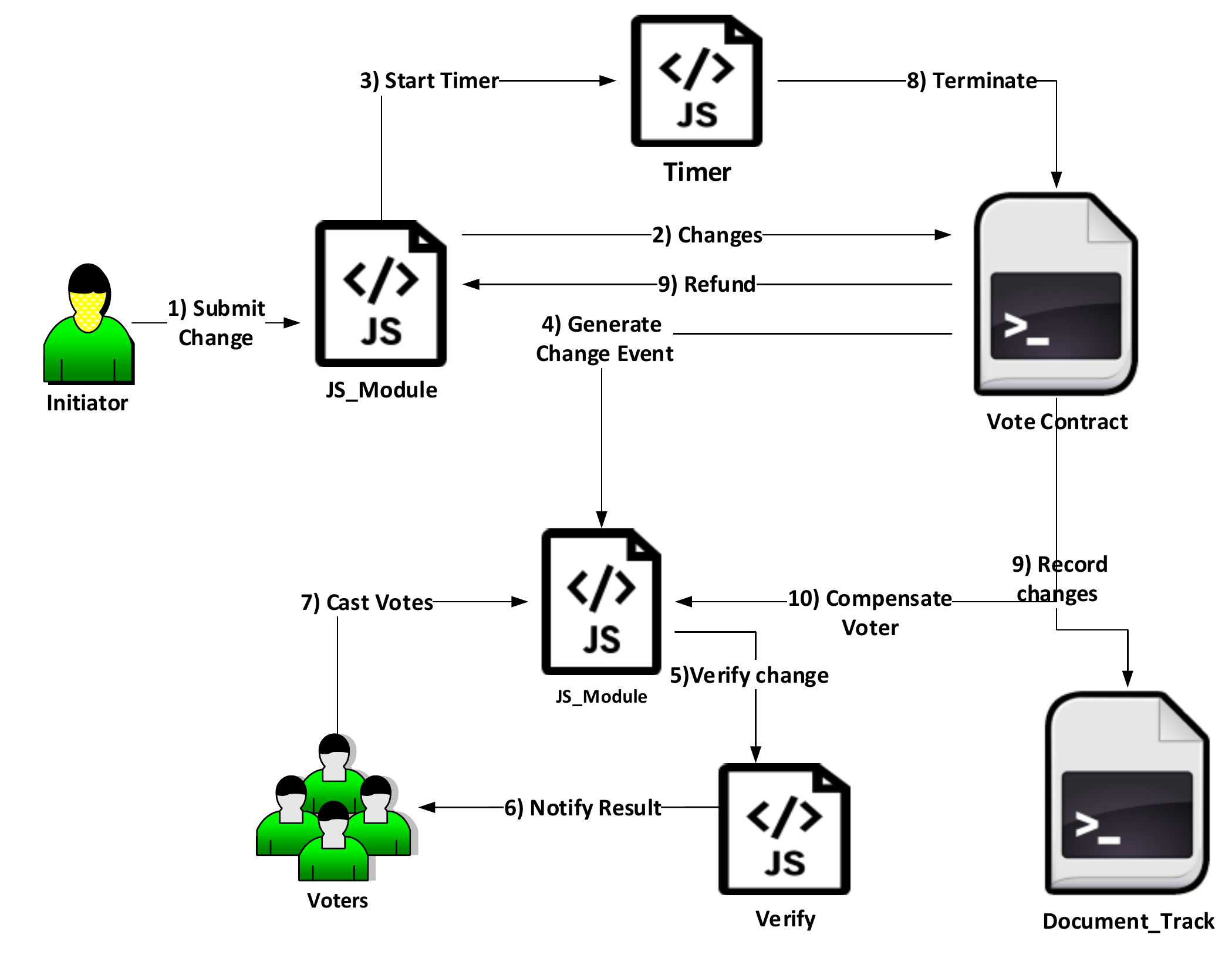}
 \caption{Voting procedure for a Document change.}
 \label{fig:vote}
\end{figure}

The overall view of the voting process is described in figure~\ref{fig:vote}. The voting process starts when the initiator submits a change to the Vote contract. The initiator client triggers a timer to initiate the voting phase of the newly submitted change. The vote contract generates an event which indicates the commencement of the voting phase for the submitted change. The Event listener module in the client applications reads the newly generated vote event. The client application verifies if it is a stakeholder in the current document change event. It then calls the verification script residing within the cloud along with the links to the current and previous versions of the file and the file hashes.  The call to verification process occurs in every node based on the voting protocol policy. If the verification script returns as true then the client application notifies the user of the result. The 
%changed
client application then casts votes on the decision to accept or reject the changes. The vote contract on receiving the vote from a client records the user decision. The timer module will terminate the vote contract at the end of the voting period. The vote contract counts both for and against votes and rejects the change if the majority have voted against the change. If the change is accepted, the deposit by the initiator is refunded back. If the change is rejected then the deposit is divided among the participants of the voting phase.  This way we incentivize truth telling by the participants and  reward participants for  catching errors.

In our current implementation of \sysname, we have implemented two types voting protocols.
\subsection{Majority voting} 
In majority voting, all the clients/users who have a stake in a document vote on a change to that document. The decision of accepting or rejecting the change is based on a simple majority. If the majority of the users votes against the change, the change is rejected, else the change is accepted. The disadvantage of this voting scheme is that it requires every user who has access to the document to vote. This policy is ideal if the number of users of a document is small(e.g., less than 5). For larger number of users, requiring all the stakeholders to vote for every change is expensive. \sysname implements simple majority voting when the number of stakeholders for the document is less than 5.

\subsection{Randomized Threshold Voting} 
Every client voting for each and every change is not efficient for systems that contain a large number of changes and users. For such scenarios, we propose randomized threshold voting. In randomized threshold voting, the contract requires that a minimum percentage of votes to accept or reject the change. Suppose the document has $n$ users, to accept or reject a change, the vote contract threshold is $s$. To ensure that each voting phase for a change receives $s$ votes, the contract tries to get expectedly $t$ votes for $t>s$. The threshold $t$ ensures that the minimum amount of for or against votes $s$ are received for each change. 

To determine whether to take part in change voting phase, each client generates a random number based on the formula: 
\[Ks = Hash(Bno,ETxt,Diff,Glim,Addr) \bmod n\]
In the formula, Ks is the random number generated by the client by hashing Bno - the current block number, Etext - the encrypted text in the change event, Diff - the current gas limit and the Addr - the initiator's address.
If the generated number is below the threshold number $t$ set by the vote contract (i.e. $Ks<t$), the client votes based on the result of the verification script. Once a vote is submitted, the vote contract generates the random number for each vote in a similar manner and verifies that the submitted vote is legitimate.

In this technique, the voting for the change is based on secure pseudo-random numbers; and it is not feasible to know which clients vote on which changes since the inputs to the hash function differs for each vote almost in a random manner. 
At the end of a voting period, if the vote contract finds that the total number of votes is below the threshold $s$, the vote contract restart the voting process. The probability of a restart event can be bounded as discussed in Section~\ref{sec:restart-analysis}. If after predefined maximum number of restarts, the required number of votes are not received, the change is rejected and the deposit is refunded to the initiator of the current change.  As we discuss  below, we can set the system parameters $t$ and $s$ in such a way that this is very unlikely. 

\subsubsection{Randomized Voting Analysis} \label{sec:restart-analysis} 
In the randomized  voting  based  verification, we  randomly choose  users  to vote for  or against the change. For security purposes, we may want to randomly choose  at least $s$ users out of $n$ users available for voting. Since the process is random, we may set the  probability that a user is randomly chosen as $\frac{t}{n}$ where $t>s$. Given this, we can analyze, the probability that a given voting phase fail to get at least $s$ votes. Before we do our analysis, we use the following Chernoff-Hoeffding bound result.

\begin{theorem}\label{thm:chernoff} \cite{paper31}
Let $X=\sum_{i=1}^n X_i$ where $X_i$, $i \in [0..n]$, are independently distributed in  $[0,1]$. Then
$Pr[X< E[X]-a] \leq e^{\frac{-2a^2}{n}}$
\end{theorem}

Now using the theorem~\ref{thm:chernoff}, we can prove the following theorem:
\begin{theorem}\label{thm:bound}
Given  $n$ users that can vote for a submitted change, for a randomized voting process that chooses a user with probability $\frac{t}{n}$ where $t>s$, and the probability $p_f$ that a chosen user does not vote due to a failure,  the probability  that  total number of users voted $V$ is less than $s$ is bounded as:
\[ 
   Pr\left[V < s\right] \leq e^{-\frac{2(t-s-n.p_f)^2}{n}}
\]
\end{theorem}
The proof of theorem~\ref{thm:bound} is given in the appendix ~\ref{prf:proof1}. We would like to the stress that $p_f$ value can be adjusted for scenarios where some users are not online and/or do not want to vote for various reasons.

\begin{figure}
\includegraphics[width=6cm,height=4cm,scale=1]{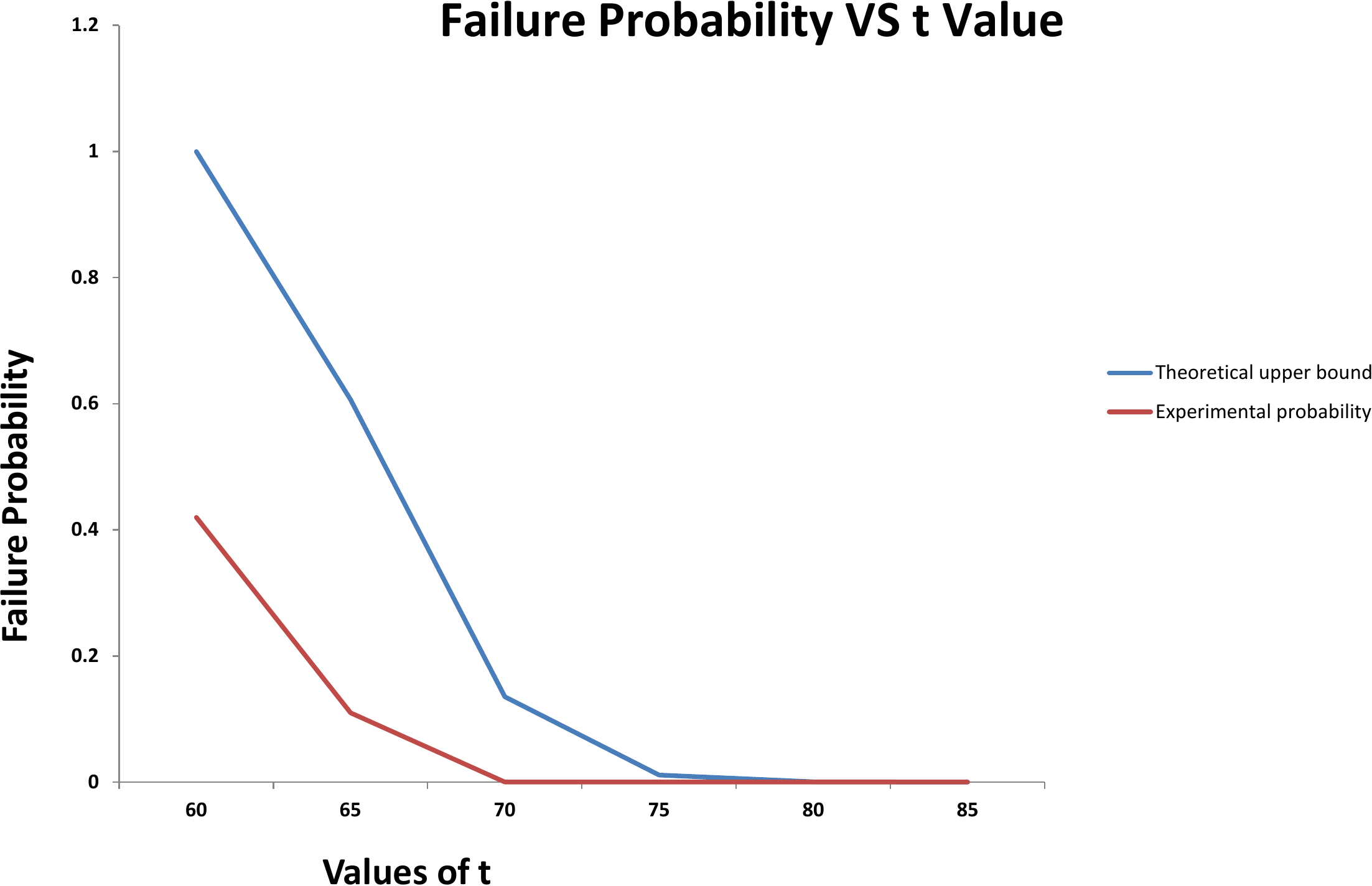}
\caption{Failure Probability}
\label{fig:restart}
\vspace{-5mm}
\end{figure}

For varying $t$ values, ($p_f=0.0$) figure~\ref{fig:restart} reports the estimated versus theoretical probability of failure for the case where $n=100$, and $s=60$. 
To estimate the failure rate in  smart contract voting, we run the  voting protocol  100 times and count the number of cases where one round voting failed to get at least $s$ votes. We use this count to estimate the probability of failure. As figure~\ref{fig:restart} shows when $t>70$, we  do not observe any failures even  though the theoretical upper bound is non-zero. 

Of course the next question is how can we set the  $t$ and $s$ values in practice so that we do not  get less than $s$ votes for each voting period.  To decide on the optimal values for $t$ and $s$, we need to consider the total cost of voting at reach round based on the failure probability (i.e., the probability that the total votes are less than $s$ ($p_t=Pr\left[V < s\right]$) for given failure probability $p_f$ value).  Below, we  show the  relationship between  $t,s,n$ and the expected cost of randomized voting process that continues until it gets at least  $s$ votes.
%As our experiments indicate a  total cost of  one round voting is a linear function  of $t$.
\begin{theorem} \label{thm:cost}
Let failure probability be $p_t$ where one round of voting gets less than $s$ votes given that each of the existing $n$ users votes with probability $\frac{t}{n}$.  Then the expected  cost  of  voting process $E[C_V]$ can be given as follows:
\[ E[C_V]= \frac{c.t+c_1}{1-p_t} \leq \frac{c.t+c_1}{1-e^{-\frac{2*(t-s)^2}{n}}}\]
for some system dependent constants $c$ and $c_1$.
\end{theorem}

Proof of the above theorem~\ref{thm:cost} is given in appendix~\ref{prf:proof2}. We can use the  theorem~\ref{thm:cost} to find the optimal value for $t$ given $s$ and $n$ based on the empirical constants  $c$ and $c_1$.  

 \section{System Analysis}\label{sec:sysanalysis}
 In this section, we analyze the security and privacy aspects of our \sysname system. Specifically, we discuss how \sysname system handles attacks from the two type of adversaries discussed in section ~\ref{sec:threat}.
 
 \subsection{Security Analysis}
 An external adversary can try to attack the current system by submitting an invalid change request for a particular document ID. \sysname contract would stop any such attempts by enforcing access control policies on documents. 
The Document Track contract will accept only those change request from users who has been granted access by the owner of the document. All other change requests are simply rejected by the contract. The Document track also penalizes the external adversary by withholding the deposit amount for the change for the attack attempt. An external adversary can mount a replay attack by using an earlier change request signature. Document track prevents this attack by keeping track of the latest change timestamp for a particular document. Any message carrying timestamp less than the latest timestamp for that document is ignored. 
 
An internal user can be the owner of the document or one of the users who has been granted access to the document by the owner. An internal adversary who is not the owner of the document can try and corrupt the data provenance trails by submitting defective changes. Since \sysname system requires each of the changes to be approved by a minimum number of users, this attack from the internal adversary succeeds only if he/she can control more than half of the total number of users allowed for the document. The randomized threshold voting further ensures that the adversary cannot know in advance which among all the voters can take part in the voting for a particular change, making it difficult to mount the attack. The internal adversary who is an owner can corrupt the system if he/she colludes with other stakeholders and votes for the change. The owner is the only user who can grant access, the system can be at a disadvantage if the owner selects a group of users who are loyal to him and corrupt the provenance trail. 
Although this type of attack may be successful, it still leaves a traceable trail on the blockchain that could be used to detect the attack.

 \subsection{Privacy Analysis}
 The privacy protection for the provenance data trail is achieved by the use of hashing and encryption. An external user can infer only the document id and the number of changes that are made to a particular document id by looking at the event logs. Each change event encrypts the payload of the event so that all an external adversary could get is the document id, the cipher text, and the signature. The link to the cloud location where the actual file is encrypted. The other information that an external user can deduce from watching the contract transaction trails are to see which users are associated with a particular document id. This information is deducible by observing iterations of the voting contract. The Ethereum platform provides anonymization of users through the use of random public addresses. The users of \sysname does not reveal their identity in the environment instead uses public addresses to perform operations in the system. In \sysname, only the file owner will have the knowledge of identity of user of the document. An adversary observing multiple voting iterations could at most deduce the public addresses associated with each document.  
 % added section about concurrency
\subsection{Concurrency}
In \sysname system, each change is the document is represented as a separate record.In the current system, we take each changes as a standalone change and does not allow multiple outstanding changes for the same document. This can be restrictive in certain use cases. The system could be modified to accept non conflicting changes in different parts of the tracked document. The system could accept changes to the document as long as they are non -conflicting there by increasing the concurrency . The above modification may involve adding an extra step to the verification script to check for non conflicting changes.    
\section{Experimental Evaluation}\label{sec:eval}
To evaluate the \sysname system, we have tested it on two real life scenarios and calculated the average cost for each of the individual operations of the smart contract. In both of the scenarios, we have found that \sysname system performs at a constant cost for individual operations and within a reasonable overhead. We provide the details of these two cases below. 

For both use cases, we have used the following evaluation setup: the client applications implemented using Meteor JS ran in a laptop(Core i7 2.4GHZ) and a desktop computer(Core i7 3.40GHZ) running Ubuntu 16.04.2 LTS. The smart contracts developed using Solidity language ran on Etheruem Ropsten Testnet. For both of the scenarios, we simulated the tests for a setting where we have 100 users for each of the document/data file. For the cloud based storage, we used Google Drive and the verification scripts were developed using Google AppScript.

\subsection{Clinical Drug trial}  
In this use case, we consider the scenario of a clinical trial~\cite{paper15} of an experimental drug. In phase 3 of the drug trial process, the drug is tested with a patient count of 300 -1000 patients. The objective of the trial is to find the side effects of varying dosages on the patients. The drug trials may be conducted by various doctors in various locations and each of the results are recorded in a common document.
\begin{figure}
\includegraphics[width=6cm,height=4cm]{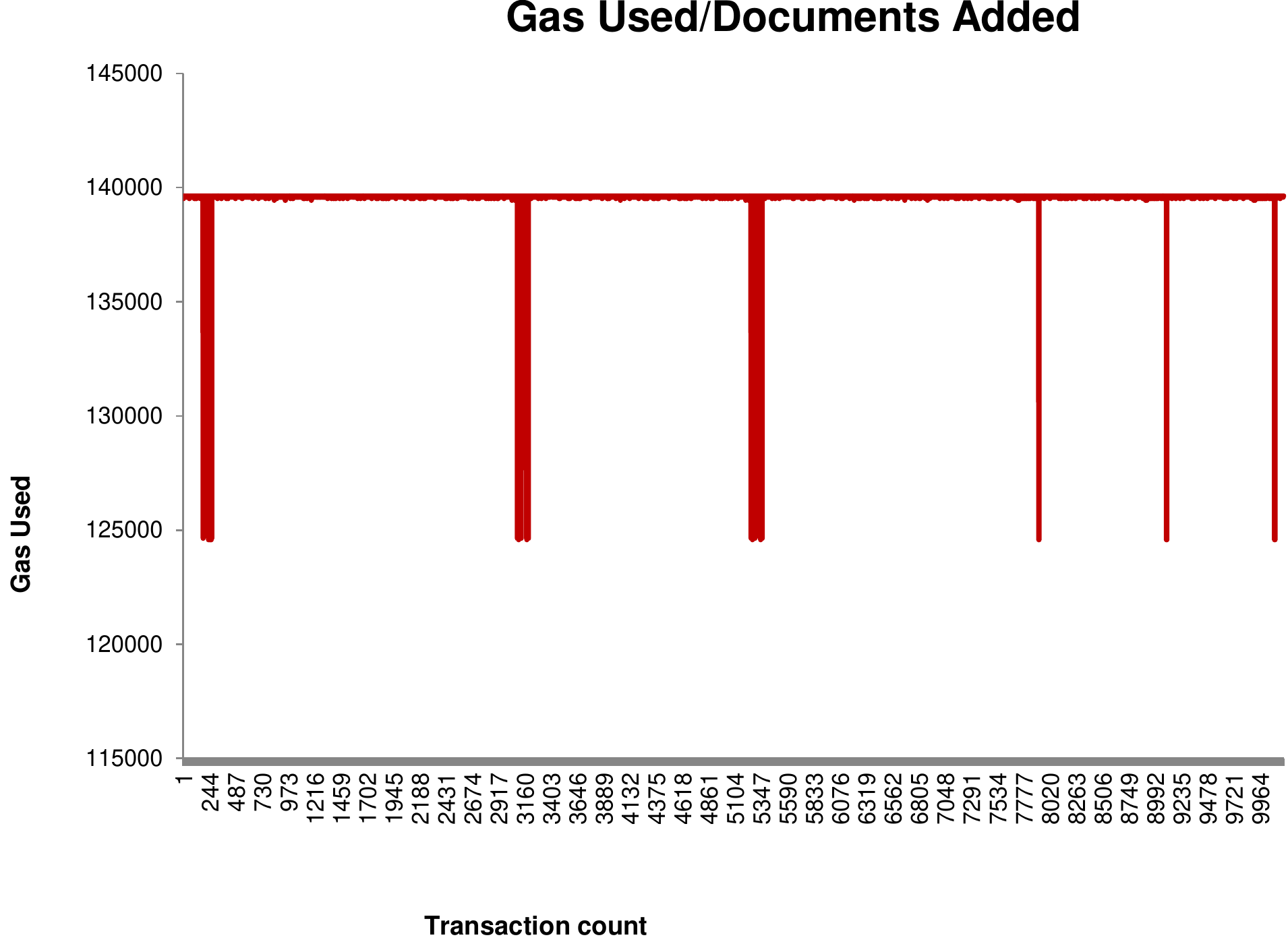}
\caption{Cost distribution of adding documents.}
\label{fig:drug-gas-document-addition}
\end{figure}
Each of the experiment group updates the same document every month for a twelve month period.  In the research setting, some of the patients may show adverse reaction to the drug. Researchers with a vested interest may try to remove those records that would show the side effects of the drug; and successive iterations of the same document will be missing records that would adversely affect the trustworthiness of the trails. To avoid the omission of records, the verification process for each of the change iterations of the document should ensure that the original patient set is maintained \footnote{ In our experiments, we chose only this constraint for automatic verification process. Other constraints could be added for different scenarios.}.

\subsubsection{Add Document}
The Add Document function is used to add a document to the system for the purpose of maintaining its provenance. The owner of the document could be the head physician who initiates the whole process. The owner generates the initial form of the file that includes the entire initial set of patient details and initial drug dosage and adds it to the contract. The AddDocument functions generates a unique document id for each file added. Figure ~\ref{fig:drug-gas-document-addition} shows the gas cost per file added for the contract. For the drug trials scenario, we need to add only a single file. The average gas cost per file added is 139552.  Please note that in our setting, \sysname keeps \textit{fixed size provenance records irrespective of the original data file size}.

\begin{figure}
\includegraphics[width=6cm,height=4cm,scale=1]{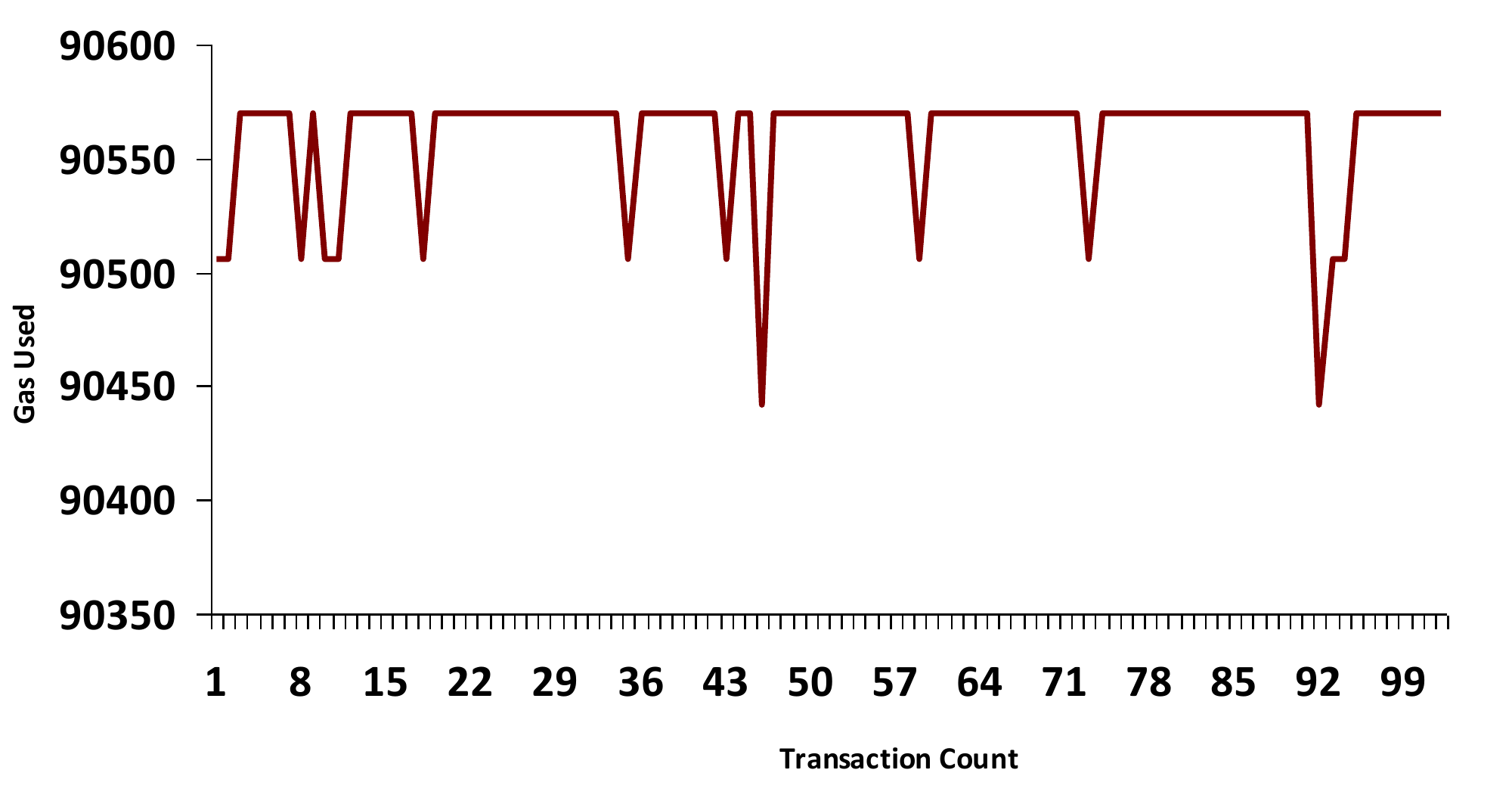}
\caption{Gas Used for Each User Added.}
\label{fig:gas-used-per-added-user}
\end{figure}
\subsubsection{Add User}
The Add user function deals with the granting access to users for a document. The user who creates a particular document is recorded as the owner of the contract. Access to a particular contract can only be granted by the owner of that contract. Figure \ref{fig:gas-used-per-added-user} gives the gas used per user added for one document where each transaction is the addition of a new user. The average gas used per transactions is 90559. The user details are stored as the hash of the user address. The spikes in the figure  \ref{fig:gas-used-per-added-user} represent the difference in the hashing requirements for the inputs.

\subsubsection{Initiate Change} 
\begin{figure}
\includegraphics[width=6cm,height=4cm,scale=1]{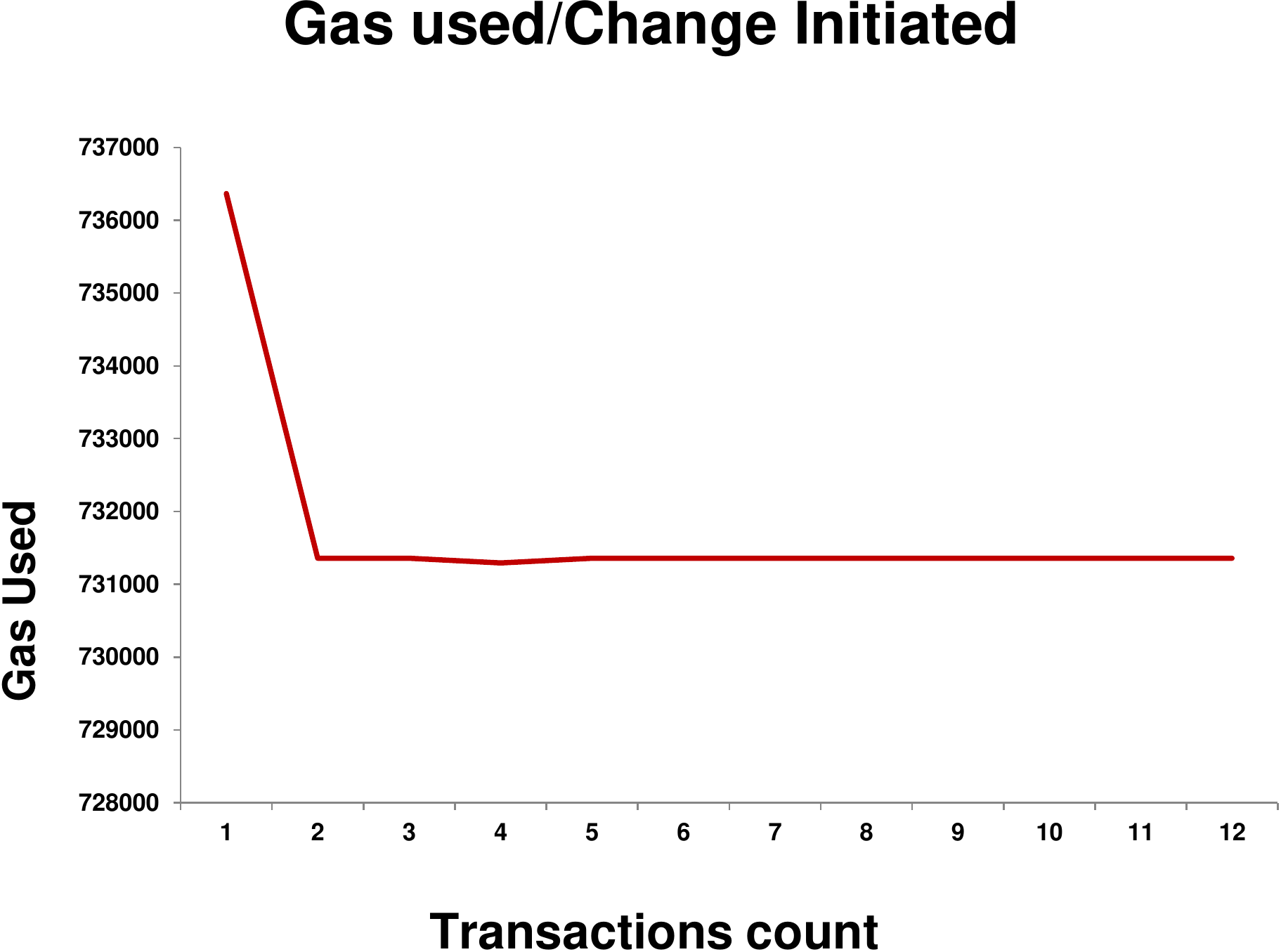}
\caption{Gas Used to vote phase initiation.}
\label{fig:vote-phase-initiation}
\end{figure}

The initiate change function deals with triggering the voting process for logging a particular change. The initiate change requires the initiator of the change to deposit an amount with the contract while calling the contract. The initiate change function is called in the current scenario at the end of every month by the doctors to record the side effects(if any) of the current dosage. The average gas used for the changes is 731768. Figure~\ref{fig:vote-phase-initiation} gives the gas distribution per initiation of voting phase for different transactions.  

\subsubsection{Voting phase} 
\begin{figure}
\includegraphics[width=6cm,height=4cm,scale=1]{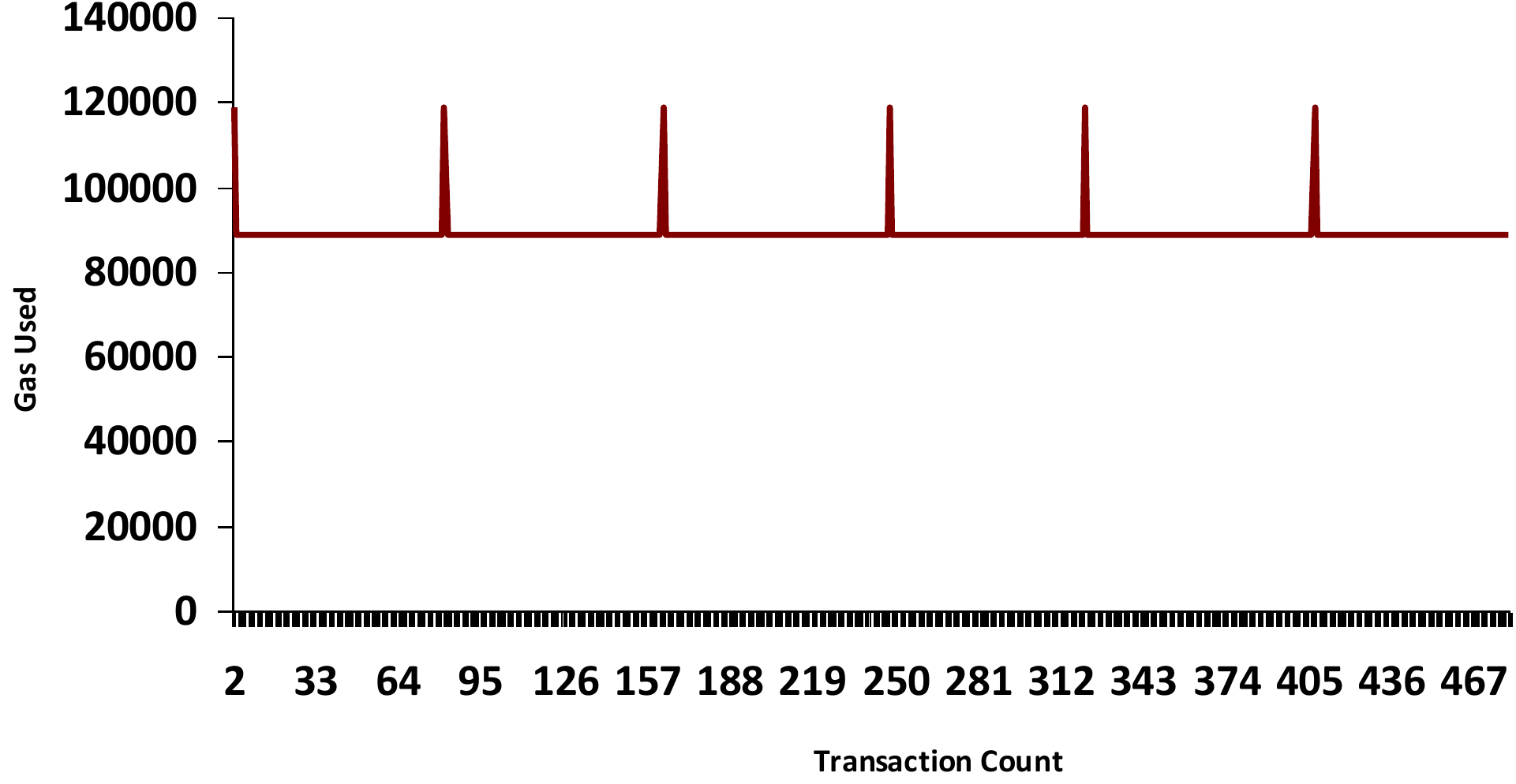}
\caption{Cost distribution of voting process.}
\label{fig:voting-phase}
\end{figure}

Once the change has been initiated, the client programs running in the voting quorum will verify the changes and cast their votes. The vote of each of the participant is recorded by the smart contract and tallied up as for and against votes. The average gas used during this process is 89176. The gas consumption for each votes is given in figure~\ref{fig:voting-phase}.
In the graph, we see that there are spikes at the beginning of each of the voting period. This is due to the initialization that occurs during the start of the voting intervals.

\begin{figure}
\includegraphics[width=6cm,height=4cm,scale=1]{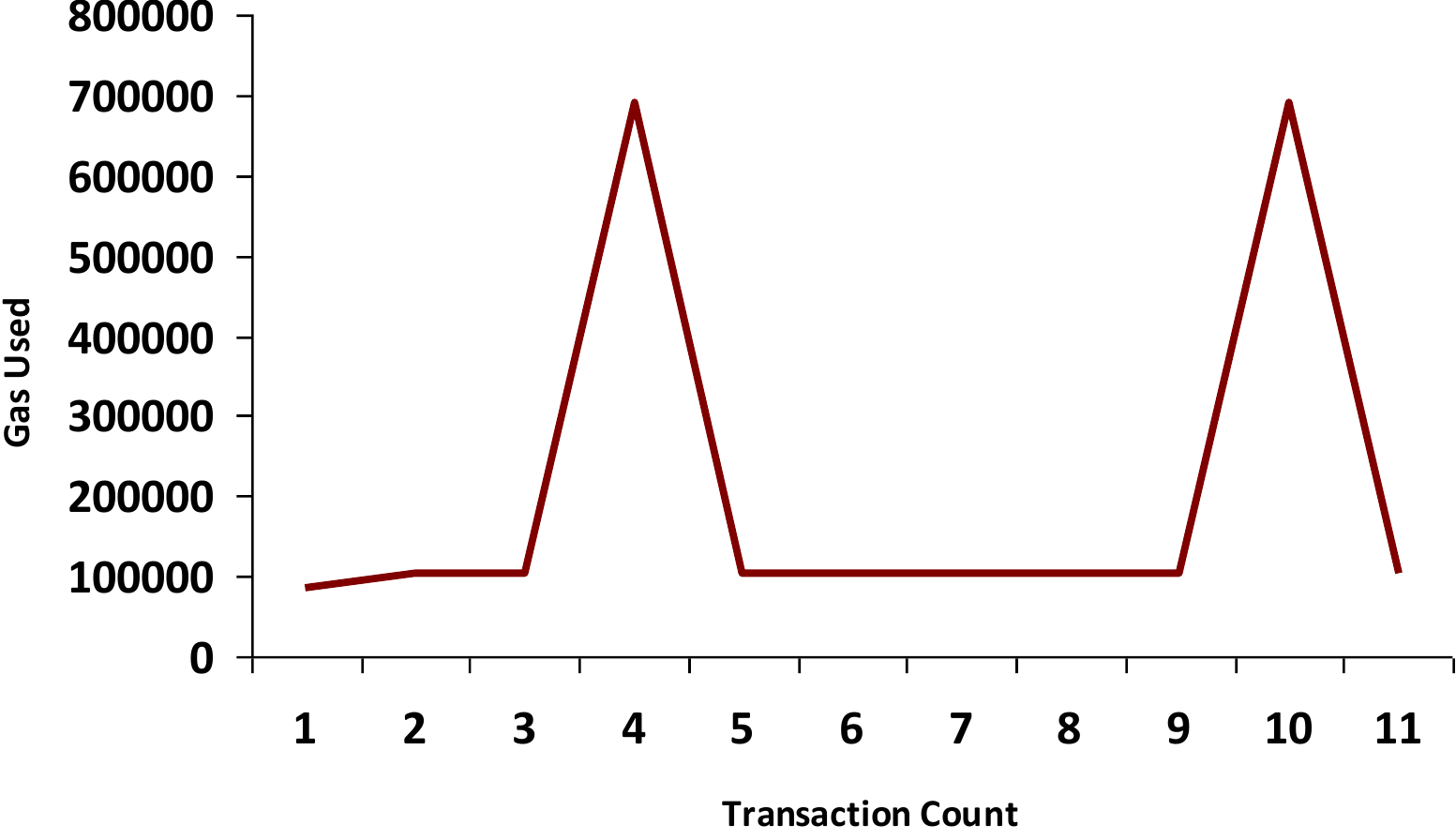}
\caption{Cost distribution of a Termination operation.}
\label{fig:termination-phase}
\end{figure}

\subsubsection{Termination:} The result of the voting process determines whether to accept or reject the changes. If the majority of the people vote against the change, then the change is rejected. On rejection of a change, the voters who verified and voted are awarded the deposit amount of the initiator. On acceptance of the change, the change is recorded in the event log of the Document track contract and the deposit is refunded to the initiator of the change. In figure~\ref{fig:termination-phase}, we can see that there are two large spikes. These are the case in which the changes are rejected after the voting process. The gas used for these are more because all the voters are awarded a part of the deposit in the case of a rejection. The average gas used for termination is 249812.

\subsubsection{Verification Script:} 
In the drug trial scenario, the verification script verifies if the same set of patients given in the original trials are maintained across the various iterations of the data collection phase. The client initiates the verification script by providing it with the link of the current file version and hashes submitted with the change. The verification script generates its own hashes and compares with the submitted hashes.The script then compares patient identification columns with the previous files to ensure that none of the original patients have been omitted from the currently submitted version. The verification script then notifies the client of the result. In the drug trials scenario, the verification script checks if all the patients records are retained in subsequent iterations.
The run times of the verification script which depends on the data file size and the verification complexity ,for data files that contain 1000 to 5000 patients, the verification run time vary between 7 secs to 31 secs.    

\subsection{Tracking Wheat Production } 
The second provenance relevant scenario that the system was adopted to is for the farming industry. The industry keeps track of the annual crop production of a given country. We implemented and tested our system based on the annual wheat production data system~\cite{paper9}. The Wheat production data file is updated quarterly and contains the quantitative details such as total production, total disappearance, and imports. 

Our provenance system implements randomized threshold voting to confirm each change to the file. The Add User, initiate vote, initiate change functions of the system use the same amount of gas as with the drug trial scenario so we do not report here. The voting phase and the termination phase saw a slight increase in the gas amount used. 

\subsubsection{Initiate vote phase:} During this phase a change to a document is submitted to the Vote Contract. The vote contract verifies the current user's access to the document and initiates the voting protocol.The average gas used is 778979.5.
\subsubsection{Vote phase:} 
\begin{figure}
\includegraphics[width=6cm,height=4cm,scale=1]{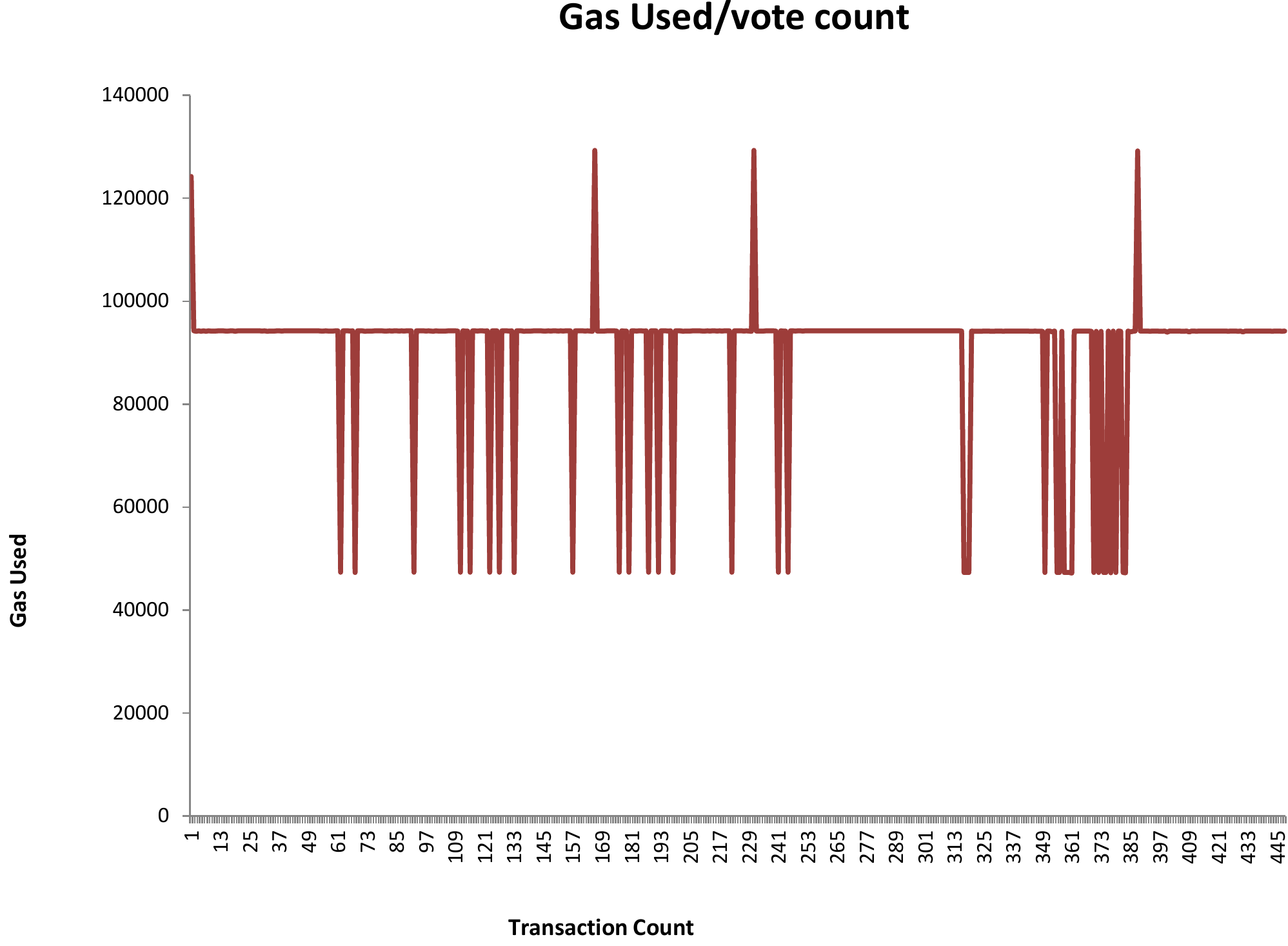}
\caption{Cost distribution of vote phase.}
\label{fig:w_vote}
\end{figure}

We implemented the threshold voting method in the system. The participating nodes generate the random number and votes if the random number falls within the defined threshold. The vote contract checks if the vote is valid before recording the vote. We simulated an internal adversary who votes out of turn. The adversary controlled nodes vote even if the random number generated does not fall within the threshold. The contract rejects these votes. The gas used for such rejections are lower than the valid votes. The average gas used is 90862.The cost distribution per votes registered is illustrated in figure \ref{fig:w_vote}. 
\subsubsection{Termination:} 
\begin{figure}
\includegraphics[width=6cm,height=4cm,scale=1]{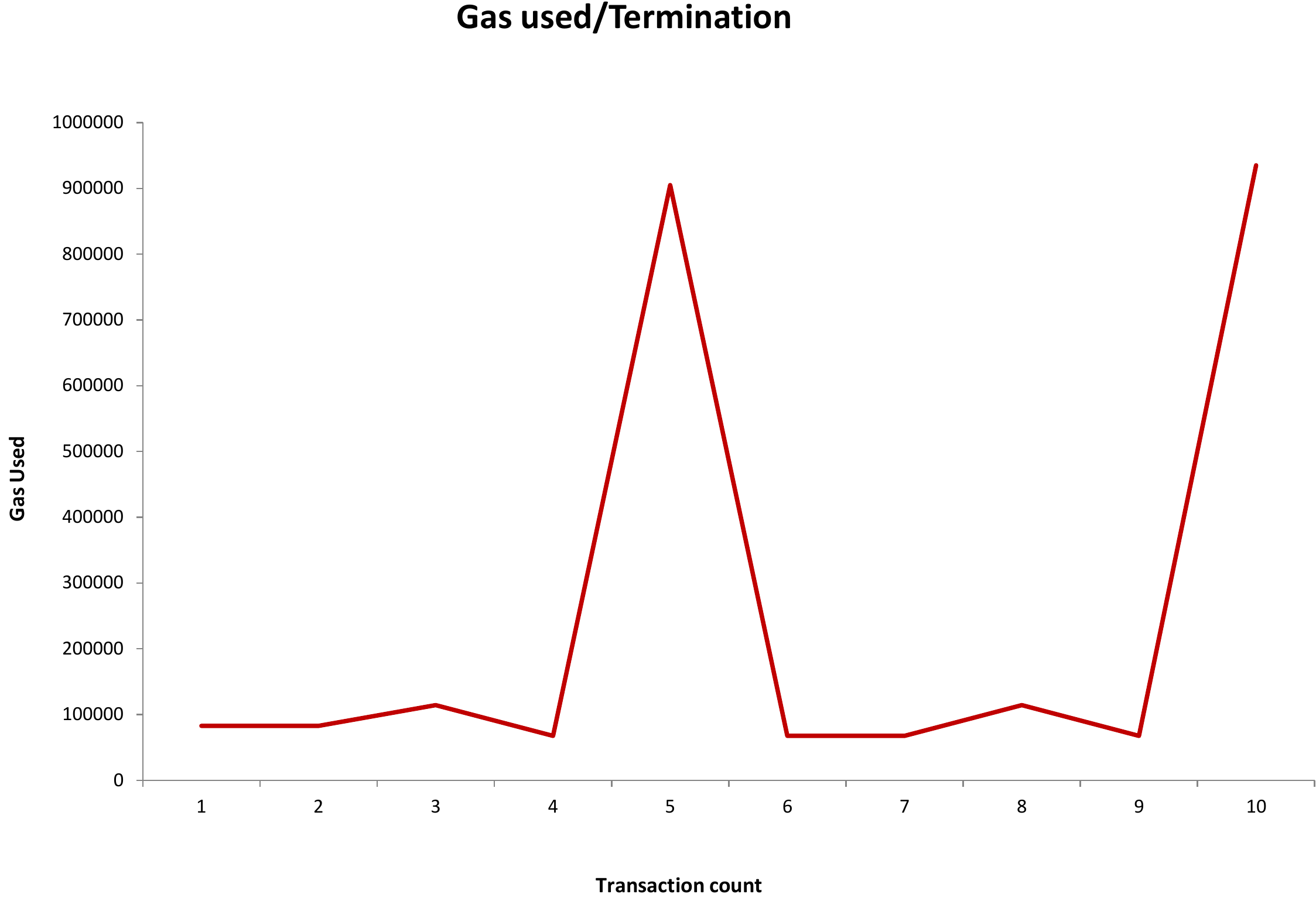}
\caption{Cost distribution of termination.}
\label{fig:w_terminate}
\end{figure}

The timer module terminates voting phase after a fixed time interval. In the threshold voting, the voting process is restarted if the threshold amount of votes are not received for a particular change. The average rate of restarts for a particular change has been found out to be 3. Our results show that restarting consumes less gas (Same amount of gas as that of submitting change) than termination of the change phase.
The cost for the termination operation is similar to the ones obtained obtained in the clinical trails experiment and not reported here.The cost distribution per votes registered is illustrated in figure \ref{fig:w_terminate}.
\subsubsection{Verification Script:} The verification script of the system compares the current files year on year data with that of the verified data.
In Wheat Tracking use case, for data files that contain 1KB to 5KB patients, the verification run time vary between 10 secs to 45 secs.

\subsection{Operation Cost}
\begin{figure}
\includegraphics[width=6cm,height=4cm,scale=1]{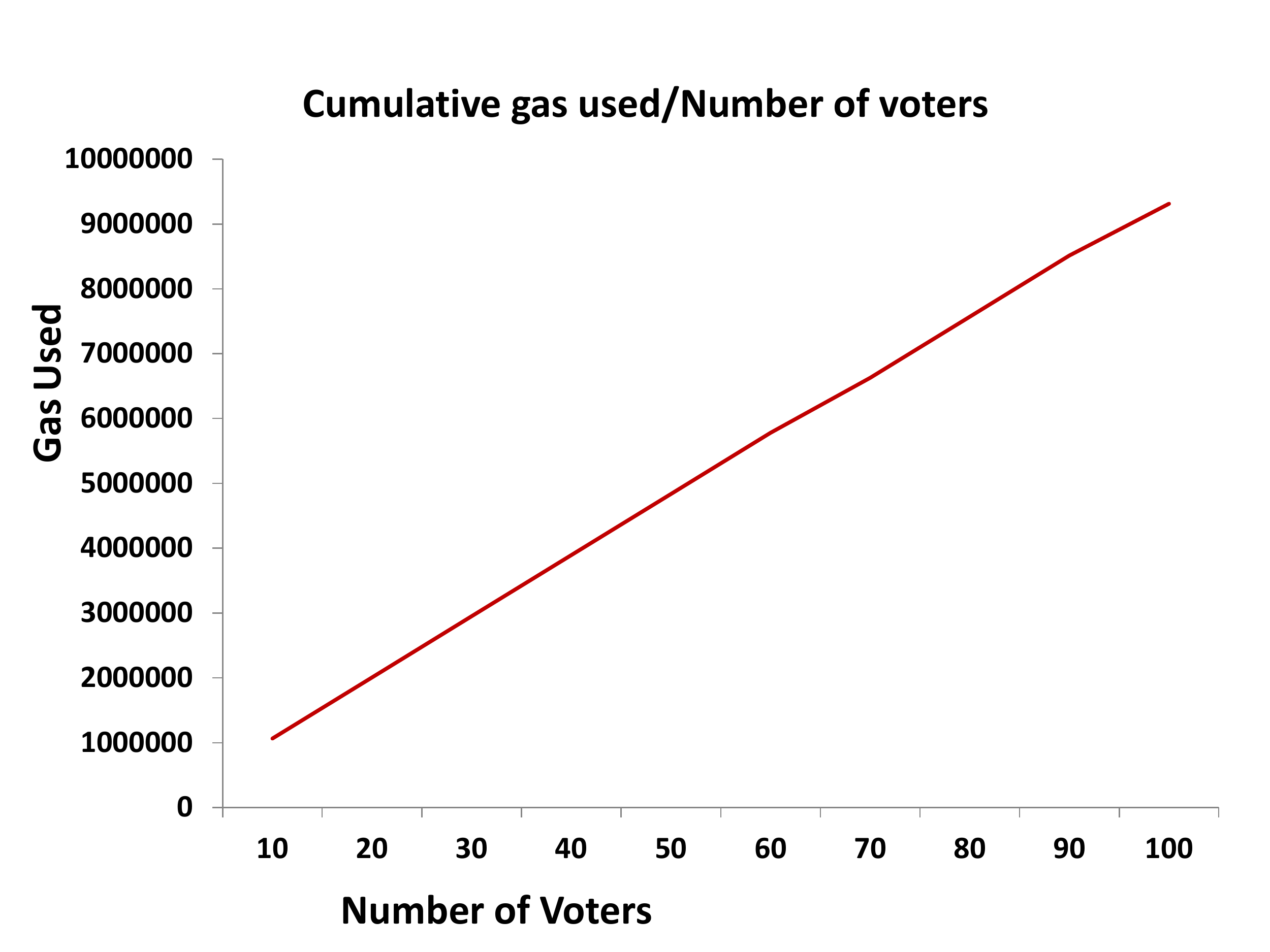}
\caption{Cumulative gas consumption for the voting process}
\label{fig:cuml-vote}
\end{figure}

By observing the system contract executions in the above scenarios, we see that for an individual functions such as add user or add document,the gas used per transaction remains almost constant. The cumulative gas used for any individual function is a near linear function (e.g., as shown in figure\ref{fig:cuml-vote} for vote function).

The gas usage is calculated at an average of 0.00000002 ethers per gas used. At the time of experimentation, a single ether cost is 90 US dollars. 
The table\ref{table:3} shows the average gas used for various operations of the \sysname system. As the results indicate most operations can be executed with relatively little cost.
\begin{table}[h!]
\centering
\begin{tabular}{||l l l ||} 
 \hline
 Operation  & Avg gas spent  & cost(USD) \\  
 \hline\hline
 Add Document & 139552  & 0.2511936\\ [1ex]
 Add User & 90559 & 0.1630062\\ [1ex]
 Initiate Change & 731351.5 & 1.3164327\\ [1ex]
 Vote & 89176.33 & 0.160517394\\ [1ex]
 Record Change	 & 249812 & 0.4496616\\ [1ex]
 \hline
\end{tabular}
\caption{Cost of operations in the \sysname System}
\label{table:3}
\end{table}
\subsection{Contract Execution Duration} 
The time taken to perform each of the operations in the system is represented in the table \ref{table:4}. We can see that all the operations take near constant time to perform. The time taken for each operation is taken as the average time taken per thousand operations.
\begin{table}[h!]
\centering
\begin{tabular}{||l l ||} 
 \hline
 Operation  & Time Taken(ms)   \\  
 \hline\hline
 Add Document & 926  \\ [1ex]
 Add User & 877 \\ [1ex]
 Initiate Change & 858 \\ [1ex]
 Vote & 829 \\ [1ex]
 Record Change	 & 950 \\ [1ex]
 \hline
\end{tabular}
\caption{Time for operations in the \sysname System}
\label{table:4}
\end{table}
The execution time of each of the above operations depends on the network speed and the speed of mining of the blocks, but in long run these times remain near constant and take less than a second in all of the usage scenarios. 
\section{Related Work}\label{sec:relatedwork}
Recently there have been several research studies that leverage blockchain as a platform for building trusted systems. Below, we summarize this work and discuss its relationship to our work.

\paragraph{Access control:} In~\cite{paper1}, authors explain the use of blockchain as a trans-organizational authentication system. The medrec system proposed in ~\cite{paper2} implements access control for medical records across medical institutions through the usage of the public blockchain.
Fairaccess system discussed in~\cite{paper17} is a decentralized access control system for the Internet of things devices using blockchain technology. In our \sysname system, we also implement access control policies, our focus is in the capturing of provenance data. 
\paragraph{Trusted Authority system:}
The legal aspect of using blockchain as a verifiable trusted source was further expanded upon by common accords group \cite{paper3} and in \cite{paper4}.
These work describe leveraging data stored in a public blockchain as a verifiable evidence in a court of law. Namecoin\cite{paper10} system uses the blockchain technology as a trusted source for the Domain Name System (DNS). 
Our \sysname system  eliminates the need for storing data on transactions by using the event logs of the smart contract to store the provenance trails. The smart contracts on top of the Ethereum platform acts as a decentralized trusted authority regarding all provenance trails stored. The provenance trails generated is also trustworthy as the decision to accept or reject a change depends on the voting protocol. \sysname therefore acts as a decentralized trust based system for data provenance.

\paragraph{Privacy preserving blockchain systems:} The DECENT system discussed in~\cite{paper11} uses the blockchain along with the smart contracts to implement key management services. It implements the idea of secret sharing to securely share keys in a public environment.The Hawk system proposed in~\cite{paper13} implements the concept of zero knowledge proofs combined with encryption to implement privacy preserving blockchain systems. The Hawk system uses two components: an on-chain component which uses smart contracts and zero-knowledge proofs to facilitates betting protocols and the off-chain components which generates zero knowledge proofs for the system. The Hawk system show how secure computations can be implemented on top of a public system such as the blockchain. 
. 
The use of secret sharing techniques for protecting sensitive information is further discussed in~\cite{paper12}. 
Compared to these works, the \sysname system utilizes encryption and hashing to preserve the privacy of the data stored in the public ethereum blockchain and secure communication channels between the smart contract and client machines to preserve the privacy. For efficiency and generalizable reasons,  verification of the captured provenance data is done off-the-chain.

 \paragraph{Security in smart contracts:} The common security vulnerabilities in the smart contracts are discussed in~\cite{paper13}. This work illustrates a number of security issues in smart contracts such as call stack bug, block hash bug, and miners withholding the addition of blocks to gain an unfair advantage. This work further discusses how to avoid these pitfalls by including additional access verification and  cryptographic primitives like encryption and hashing. 
Compared to these works, \sysname implements digital signatures to avoid malicious logging of provenance data. It uses an encrypted form of the provenance trails to avoid revealing details such as the location of the files and the user access information. \sysname further restricts the access to methods based on checks implemented on the user address.

\paragraph{Data Provenance:} 
Leveraging the blockchain as a data provenance tracker was first discussed by the Project Provenance~\cite{paper18}. In this work, blockchain transactions are used to store provenance details of food products from production to the consumer. In addition, in \cite{paper19},  the use of blockchain as provenance platform is presented as one of the four breakout cases of the blockchain platform. The use of bitcoin as a data provenance system for research scenario was further explored in~\cite{paper7}. The author suggested the idea of storing the research objectives as an encoded file in the data fields of bitcoin transactions. Compared to these works,
the \sysname system  adopts the immutability of the blockchain environment and  implements a full stack privacy preserving, verified data provenance store with access control policies. The provenance chains that are generated by the \sysname system are stored as event logs there by saving costs on storage. The system facilitates the verification of these provenance events by any authorized users. \sysname provides a platform to implement custom verification scripts suited for the application area. The system ensures privacy by using public key encryption and preserves integrity by the use of digital signatures.
%Added description about PROVchain compare and contrast
The ProvChain~\cite{paper33} system provides a data provenance system based on blockchain  technology. The ProvChain system uses monitor programs called 'hooks' to track the changes that occur in the cloud storage system and records each and generates events corresponding to the actions of the users. The user events thus recorded are then stored on the blockchain as transactions. The verification process is achieved by an external entity known as auditor. The auditor generates transaction receipts using Tieron API~\cite{paper18}. The Provchain system verifies the changes after the information is logged on to the blockchain.
The \sysname differs from Provchain by implementing automated verification scripts and rejecting the invalid changes. The change hash-chain generated by the \sysname records only the changes that are verified by the verification script. This guarantees that the changed document is always valid and prevents any chance of collusion between the auditor and the stakeholders. Another major difference of compared to Provchain is that \sysname implements incentivized voting using smart contracts to penalize the users who tries to log invalid changes to the system. The use of randomized voting reduces the centralization of the verification process. Therefore, there is no need for a physical verifier as the verification script verifies the changes before voting on the changes. The advantage of developing verification script is that a  verification script for a scenario could be reused by similar applications there by reducing the cost of development.

\section{Conclusion}\label{sec:conclude}
The \sysname is a blockchain based system that provides access control based privacy-preserving data provenance trails. In \sysname system, an authorized user can verify the changes that are made to any data file.  It also provides a proof of change with the use of digital signatures and timestamping. The system ensures that the change logs in the blockchain environment are only accessed by the authorized users with appropriate keys. The \sysname system further enhances trustworthiness of the data trails by implementing randomized voting for the change trails recorded/captured and any deviation is punished by a monetary penalty using smart contracts. The evaluation of the system based on two real life scenarios has shown that individual operations of the  system runs with acceptable cost and near constant time. 
\section{Appendix}\label{app:appendix}
\subsection{Change Log Event Format}

 change(docid,agent,$E_K$(docid,$H(X_o) $,$H(X_n) $,link,ts),OPM,$sign_k$)
  \\ where\\
  \hspace*{0.1cm}docid = unique identifier of the Document.\\
  \hspace*{0.1cm}agent = address of change initiator \\
  \hspace*{0.1cm}$E_K$ = Encrypted Text \\
   \hspace*{0.1cm}$H(X_o) $= Hash of previous file version  \\
   \hspace*{0.1cm}$H(X_n) $= Hash of new file version  \\
   \hspace*{0.1cm}link = link to cloud storage location  \\
  \hspace*{0.1cm}opm = High level representation of the OPM model  \\
  \hspace*{0.1cm}ts = latest timestamp  \\
  \hspace*{0.1cm}$sign_k$ = signature of initiator based on Encrypted text
\subsection{Proof of Theorem 5.2}
\begin{proof}\label{prf:proof1}
Let $V_i$ be the  indicator  variable that  gets the  value $1$ if  $i^{th}$ user is selected by the  randomized voting process.  
By definition, we can write $V=\sum_{i=1}^n V_i$.  Therefore, expected value of $V$ can be written as:
\[E(V)=E(\sum_{i=1}^n V_i)=n* E(V_1)=n*(\frac{t}{n}-p_f)=t-n.p_f\]
Now we  can use the theorem~\ref{thm:chernoff}, and set $a=(t-n.p_f-s)$ and $E[V]=t-n.p_f$. This concludes the proof.
\end{proof}
\subsection{Proof of Theorem 5.3}
\begin{proof}\label{prf:proof2}
For a given one round voting failure probability $p_t$ and the cost of one round voting $C_t$, we can compute the  expected cost of voting phase (i.e., voting continues till we have a round where at least $s$ users voted) as follows:
\begin{eqnarray*}
   E[C_V]&=&(1-p_t).C_t+p_t(E[C_V]+C_t)\\
         &=& C_t.+p_t.E[C_V] \\
         &=&\frac{C_t}{1-p_t} 
\end{eqnarray*}
In the above  equation we use the fact that the expected cost is $C_t$ if there is no failure. If there is failure, we pay the cost of one round of voting and then restart the voting from scratch. Solving this recursive equation  gives us the  required result. Furthermore, our empirical  analysis show that total cost of  one round voting $C_t$ is  a linear function  of $t$, and can be represented as a linear function $c.t+c_1$ for some constants $c$ and $c_1$.  So replacing $C_t=c.t+c_1$ and using the theorem~\ref{thm:bound} to bound $p_t \leq e^{-\frac{2*(t-s)^2}{n}}$ concludes our proof.  
\end{proof}

\bibliographystyle{ACM-Reference-Format}
\bibliography{sigproc.bib} 

%%% -*-BibTeX-*-
%%% Do NOT edit. File created by BibTeX with style
%%% ACM-Reference-Format-Journals [18-Jan-2012].

\begin{thebibliography}{00}

%%% ====================================================================
%%% NOTE TO THE USER: you can override these defaults by providing
%%% customized versions of any of these macros before the \bibliography
%%% command.  Each of them MUST provide its own final punctuation,
%%% except for \shownote{}, \showDOI{}, and \showURL{}.  The latter two
%%% do not use final punctuation, in order to avoid confusing it with
%%% the Web address.
%%%
%%% To suppress output of a particular field, define its macro to expand
%%% to an empty string, or better, \unskip, like this:
%%%
%%% \newcommand{\showDOI}[1]{\unskip}   % LaTeX syntax
%%%
%%% \def \showDOI #1{\unskip}           % plain TeX syntax
%%%
%%% ====================================================================

\ifx \showCODEN    \undefined \def \showCODEN     #1{\unskip}     \fi
\ifx \showDOI      \undefined \def \showDOI       #1{#1}\fi
\ifx \showISBNx    \undefined \def \showISBNx     #1{\unskip}     \fi
\ifx \showISBNxiii \undefined \def \showISBNxiii  #1{\unskip}     \fi
\ifx \showISSN     \undefined \def \showISSN      #1{\unskip}     \fi
\ifx \showLCCN     \undefined \def \showLCCN      #1{\unskip}     \fi
\ifx \shownote     \undefined \def \shownote      #1{#1}          \fi
\ifx \showarticletitle \undefined \def \showarticletitle #1{#1}   \fi
\ifx \showURL      \undefined \def \showURL       {\relax}        \fi
% The following commands are used for tagged output and should be
% invisible to TeX
\providecommand\bibfield[2]{#2}
\providecommand\bibinfo[2]{#2}
\providecommand\natexlab[1]{#1}
\providecommand\showeprint[2][]{arXiv:#2}

\bibitem[\protect\citeauthoryear{Ariel~Ekblaw}{Ariel~Ekblaw}{2016}]%
        {paper2}
\bibfield{author}{\bibinfo{person}{Thiago Vieira Andrew~Lippman Ariel~Ekblaw,
  Asaf~Azaria}.} \bibinfo{year}{2016}\natexlab{}.
\newblock \showarticletitle{MedRec: Medical Data Management on the Blockchain}.
\newblock  (\bibinfo{year}{2016}).
\newblock
\newblock
\shownote{version: 57e013615dbf3f3300152554.}


\bibitem[\protect\citeauthoryear{Bollier}{Bollier}{2015}]%
        {paper3}
\bibfield{author}{\bibinfo{person}{David Bollier}.}
  \bibinfo{year}{2015}\natexlab{}.
\newblock \bibinfo{title}{Reinventing Law for the Commons}.
\newblock   (\bibinfo{year}{2015}).
\newblock
\showURL{%
\url{http://www.commonaccord.org/}}


\bibitem[\protect\citeauthoryear{Brakerski, Katz, Segev, and
  Yerukhimovich}{Brakerski et~al\mbox{.}}{2011}]%
        {paper28}
\bibfield{author}{\bibinfo{person}{Zvika Brakerski}, \bibinfo{person}{Jonathan
  Katz}, \bibinfo{person}{Gil Segev}, {and} \bibinfo{person}{Arkady
  Yerukhimovich}.} \bibinfo{year}{2011}\natexlab{}.
\newblock \showarticletitle{Limits on the Power of Zero-Knowledge Proofs in
  Cryptographic Constructions}. In \bibinfo{booktitle}{{\em Theory of
  Cryptography - 8th Theory of Cryptography Conference, {TCC} 2011, Providence,
  RI, USA, March 28-30, 2011. Proceedings}}. \bibinfo{pages}{559--578}.
\newblock
\showDOI{%
\url{https://doi.org/10.1007/978-3-642-19571-6_34}}


\bibitem[\protect\citeauthoryear{Brown}{Brown}{2015}]%
        {paper26}
\bibfield{author}{\bibinfo{person}{Jonathan Brown}.}
  \bibinfo{year}{2015}\natexlab{}.
\newblock \bibinfo{title}{Storing compressed text in Ethereum transaction
  logs}.
\newblock   (\bibinfo{year}{2015}).
\newblock
\showURL{%
\url{http://jonathanpatrick.me/blog/ethereum-compressed-text}}


\bibitem[\protect\citeauthoryear{Buterin}{Buterin}{2015}]%
        {paper8}
\bibfield{author}{\bibinfo{person}{Vitalik Buterin}.}
  \bibinfo{year}{2015}\natexlab{}.
\newblock \showarticletitle{A Next-Generation Smart Contract and Decentralized
  Application Platform}.
\newblock  (\bibinfo{year}{2015}).
\newblock
\newblock
\shownote{September.}


\bibitem[\protect\citeauthoryear{Clark, Ciccarese, and Goble}{Clark
  et~al\mbox{.}}{2013}]%
        {paper30}
\bibfield{author}{\bibinfo{person}{Tim Clark}, \bibinfo{person}{Paolo
  Ciccarese}, {and} \bibinfo{person}{Carole~A. Goble}.}
  \bibinfo{year}{2013}\natexlab{}.
\newblock \showarticletitle{Micropublications: a Semantic Model for Claims,
  Evidence, Arguments and Annotations in Biomedical Communications}.
\newblock \bibinfo{journal}{{\em CoRR\/}}  \bibinfo{volume}{abs/1305.3506}
  (\bibinfo{year}{2013}).
\newblock
\showURL{%
\url{http://arxiv.org/abs/1305.3506}}


\bibitem[\protect\citeauthoryear{Durham}{Durham}{2010}]%
        {paper10}
\bibfield{author}{\bibinfo{person}{Vincent Durham}.}
  \bibinfo{year}{2010}\natexlab{}.
\newblock \bibinfo{booktitle}{{\em NAMECOIN}}.
\newblock
\newblock
\shownote{\url{https://namecoin.org/}.}


\bibitem[\protect\citeauthoryear{Economist}{Economist}{2016}]%
        {paper7}
\bibfield{author}{\bibinfo{person}{The Economist}.}
  \bibinfo{year}{2016}\natexlab{}.
\newblock \showarticletitle{Better with bitcoin}.
\newblock  (\bibinfo{year}{2016}).
\newblock
\newblock
\shownote{http://www.economist.com/news/science-and-technology/21699099-blockchain-technology-could-improve-reliability-medical-trials-better.}


\bibitem[\protect\citeauthoryear{Food and Administration}{Food and
  Administration}{2017}]%
        {paper15}
\bibfield{author}{\bibinfo{person}{US Food} {and} \bibinfo{person}{Drug
  Administration}.} \bibinfo{year}{2017}\natexlab{}.
\newblock \bibinfo{booktitle}{{\em Clinical Research}}.
\newblock
\newblock
\shownote{\url{https://www.fda.gov/ForPatients/Approvals/Drugs/ucm405622.htm}.}


\bibitem[\protect\citeauthoryear{George~SL}{George~SL}{2015}]%
        {paper23}
\bibfield{author}{\bibinfo{person}{Buyse~M George~SL}.}
  \bibinfo{year}{2015}\natexlab{}.
\newblock \showarticletitle{Data fraud in clinical trials}.
\newblock \bibinfo{journal}{{\em PMC\/}} \bibinfo{volume}{5},
  \bibinfo{number}{2} (\bibinfo{year}{2015}), \bibinfo{pages}{161--173}.
\newblock
\showDOI{%
\url{https://doi.org/10.4155/cli}}


\bibitem[\protect\citeauthoryear{{G}ipp, {K}osti, and {B}reitinger}{{G}ipp
  et~al\mbox{.}}{2016}]%
        {paper4}
\bibfield{author}{\bibinfo{person}{{B}ela {G}ipp}, \bibinfo{person}{{J}agrut
  {K}osti}, {and} \bibinfo{person}{{C}orinna {B}reitinger}.}
  \bibinfo{year}{2016}\natexlab{}.
\newblock \showarticletitle{{Securing Video Integrity Using Decentralized
  Trusted Timestamping on the Blockchain}}. In \bibinfo{booktitle}{{\em
  {P}roceedings of the 10th {M}editerranean {C}onference on {I}nformation
  {S}ystems ({MCIS})}}. \bibinfo{address}{Paphos, Cyprus}.
\newblock


\bibitem[\protect\citeauthoryear{Goldreich}{Goldreich}{2004}]%
        {paper32}
\bibfield{author}{\bibinfo{person}{Oded Goldreich}.}
  \bibinfo{year}{2004}\natexlab{}.
\newblock \bibinfo{booktitle}{{\em Foundations of Cryptography: Volume 2, Basic
  Applications}}.
\newblock \bibinfo{publisher}{Cambridge University Press},
  \bibinfo{address}{New York, NY, USA}.
\newblock
\showISBNx{0521830842}


\bibitem[\protect\citeauthoryear{Google}{Google}{2017}]%
        {paper24}
\bibfield{author}{\bibinfo{person}{Google}.} \bibinfo{year}{2017}\natexlab{}.
\newblock \bibinfo{booktitle}{{\em Google Appscript}}.
\newblock
\newblock
\shownote{\url{https://developers.google.com/apps-script/}.}


\bibitem[\protect\citeauthoryear{Greenspan}{Greenspan}{2016}]%
        {paper19}
\bibfield{author}{\bibinfo{person}{Gideon Greenspan}.}
  \bibinfo{year}{2016}\natexlab{}.
\newblock \bibinfo{title}{Four Genuine Blockchain Use Cases}.
\newblock   (\bibinfo{date}{May} \bibinfo{year}{2016}).
\newblock
\showURL{%
\url{http://www.coindesk.com/four-genuine-blockchain-use-cases/}}


\bibitem[\protect\citeauthoryear{Hills and Ramapriyan}{Hills and
  Ramapriyan}{2015}]%
        {paper5}
\bibfield{author}{\bibinfo{person}{R.~R. Downs R. Duerr J. C. Goldstein M.
  A.~Parsons Hills, D.~J.} {and} \bibinfo{person}{H.~K. Ramapriyan}.}
  \bibinfo{year}{2015}\natexlab{}.
\newblock \showarticletitle{The importance of data set provenance for science}.
\newblock  (\bibinfo{year}{2015}).
\newblock
\newblock
\shownote{version: doi:10.1029/2015EO040557.}


\bibitem[\protect\citeauthoryear{Jagomägis, Laud, and Pankova}{Jagomägis
  et~al\mbox{.}}{2015}]%
        {paper12}
\bibfield{author}{\bibinfo{person}{Roman Jagomägis}, \bibinfo{person}{Peeter
  Laud}, {and} \bibinfo{person}{Alisa Pankova}.}
  \bibinfo{year}{2015}\natexlab{}.
\newblock \bibinfo{title}{Preprocessing-Based Verification of Multiparty
  Protocols with Honest Majority}.
\newblock \bibinfo{howpublished}{Cryptology ePrint Archive, Report 2015/674}.
  (\bibinfo{year}{2015}).
\newblock
\newblock
\shownote{\url{http://eprint.iacr.org/2015/674}.}


\bibitem[\protect\citeauthoryear{Jason and Yuichi}{Jason and Yuichi}{2015}]%
        {paper1}
\bibfield{author}{\bibinfo{person}{Cruz Jason, Paul} {and}
  \bibinfo{person}{Kaji Yuichi}.} \bibinfo{year}{2015}\natexlab{}.
\newblock \showarticletitle{The Bitcoin Network as Platform for
  Trans-Organizational Attribute Authentication}.
\newblock \bibinfo{journal}{{\em IPSJ SIG Notes\/}} \bibinfo{volume}{2015},
  \bibinfo{number}{12} (\bibinfo{date}{feb} \bibinfo{year}{2015}),
  \bibinfo{pages}{1--6}.
\newblock
\showISSN{09196072}
\showURL{%
\url{http://ci.nii.ac.jp/naid/110009877764/en/}}


\bibitem[\protect\citeauthoryear{Kevin~Delmolino and Shi}{Kevin~Delmolino and
  Shi}{2015}]%
        {paper13}
\bibfield{author}{\bibinfo{person}{Ahmed Kosba Andrew~Miller Kevin~Delmolino,
  Mitchell~Arnett} {and} \bibinfo{person}{Elaine Shi}.}
  \bibinfo{year}{2015}\natexlab{}.
\newblock \bibinfo{title}{Step by Step Towards Creating a Safe Smart Contract:
  Lessons and Insights from a Cryptocurrency Lab}.
\newblock \bibinfo{howpublished}{Cryptology ePrint Archive, Report 2015/460}.
  (\bibinfo{year}{2015}).
\newblock
\newblock
\shownote{\url{http://eprint.iacr.org/2015/460}.}


\bibitem[\protect\citeauthoryear{Kosba, Miller, Shi, Wen, and
  Papamanthou}{Kosba et~al\mbox{.}}{2015}]%
        {paper27}
\bibfield{author}{\bibinfo{person}{Ahmed Kosba}, \bibinfo{person}{Andrew
  Miller}, \bibinfo{person}{Elaine Shi}, \bibinfo{person}{Zikai Wen}, {and}
  \bibinfo{person}{Charalampos Papamanthou}.} \bibinfo{year}{2015}\natexlab{}.
\newblock \bibinfo{title}{Hawk: The Blockchain Model of Cryptography and
  Privacy-Preserving Smart Contracts}.
\newblock \bibinfo{howpublished}{Cryptology ePrint Archive, Report 2015/675}.
  (\bibinfo{year}{2015}).
\newblock
\newblock
\shownote{\url{http://eprint.iacr.org/2015/675}.}


\bibitem[\protect\citeauthoryear{Liang, Shetty, Tosh, Kamhoua, Kwiat, and
  Njilla}{Liang et~al\mbox{.}}{2017}]%
        {paper33}
\bibfield{author}{\bibinfo{person}{Xueping Liang}, \bibinfo{person}{Sachin
  Shetty}, \bibinfo{person}{Deepak Tosh}, \bibinfo{person}{Charles Kamhoua},
  \bibinfo{person}{Kevin Kwiat}, {and} \bibinfo{person}{Laurent Njilla}.}
  \bibinfo{year}{2017}\natexlab{}.
\newblock \showarticletitle{ProvChain: A Blockchain-based Data Provenance
  Architecture in Cloud Environment with Enhanced Privacy and Availability}. In
  \bibinfo{booktitle}{{\em Proceedings of the 17th IEEE/ACM International
  Symposium on Cluster, Cloud and Grid Computing}} {\em
  (\bibinfo{series}{CCGrid '17})}. \bibinfo{publisher}{IEEE Press},
  \bibinfo{address}{Piscataway, NJ, USA}, \bibinfo{pages}{468--477}.
\newblock
\showISBNx{978-1-5090-6610-0}
\showDOI{%
\url{https://doi.org/10.1109/CCGRID.2017.8}}


\bibitem[\protect\citeauthoryear{Linder}{Linder}{2016}]%
        {paper11}
\bibfield{author}{\bibinfo{person}{Peter Linder}.}
  \bibinfo{year}{2016}\natexlab{}.
\newblock \bibinfo{title}{DEcryption Contract ENforcement Tool (DECENT): A
  Practical Alternative to Government Decryption Backdoors}.
\newblock \bibinfo{howpublished}{Cryptology ePrint Archive, Report 2016/245}.
  (\bibinfo{year}{2016}).
\newblock
\newblock
\shownote{\url{http://eprint.iacr.org/2016/245}.}


\bibitem[\protect\citeauthoryear{Ltd}{Ltd}{2015}]%
        {paper18}
\bibfield{author}{\bibinfo{person}{Project~Provenance Ltd}.}
  \bibinfo{year}{2015}\natexlab{}.
\newblock \bibinfo{title}{Blockchain: the solution for transparency in product
  supply chains}.
\newblock   (\bibinfo{year}{2015}).
\newblock
\showURL{%
\url{https://www.provenance.org/whitepaper}}


\bibitem[\protect\citeauthoryear{MeteorJs}{MeteorJs}{2016}]%
        {paper22}
\bibfield{author}{\bibinfo{person}{MeteorJs}.} \bibinfo{year}{2016}\natexlab{}.
\newblock   (\bibinfo{date}{May} \bibinfo{year}{2016}).
\newblock
\showURL{%
\url{https://www.meteor.com/}}


\bibitem[\protect\citeauthoryear{MongoDB}{MongoDB}{2017}]%
        {paper25}
\bibfield{author}{\bibinfo{person}{MongoDB}.} \bibinfo{year}{2017}\natexlab{}.
\newblock \bibinfo{title}{MongoDB}.
\newblock   (\bibinfo{date}{Jan.} \bibinfo{year}{2017}).
\newblock
\showURL{%
\url{https://www.mongodb.com/}}


\bibitem[\protect\citeauthoryear{Open Provenance model}{Open Provenance
  model}{2007}]%
        {paper6}
Open Provenance model \bibinfo{year}{2007}\natexlab{}.
\newblock \bibinfo{booktitle}{{\em Open Provenance Model}}.
\newblock Open Provenance model.
\newblock
\newblock
\shownote{\url{http://openprovenance.org/}.}


\bibitem[\protect\citeauthoryear{Ouaddah, Kalam, and Ouahman}{Ouaddah
  et~al\mbox{.}}{2016}]%
        {paper17}
\bibfield{author}{\bibinfo{person}{Aafaf Ouaddah}, \bibinfo{person}{Anas
  Abou~El Kalam}, {and} \bibinfo{person}{Abdellah~Ait Ouahman}.}
  \bibinfo{year}{2016}\natexlab{}.
\newblock \showarticletitle{FairAccess: a new Blockchain-based access control
  framework for the Internet of Things}.
\newblock \bibinfo{journal}{{\em Security and Communication Networks\/}}
  \bibinfo{volume}{9}, \bibinfo{number}{18} (\bibinfo{year}{2016}),
  \bibinfo{pages}{5943--5964}.
\newblock
\showDOI{%
\url{https://doi.org/10.1002/sec.1748}}


\bibitem[\protect\citeauthoryear{Pettersen, Goddard, Huang, Couch, Greenblatt,
  Meng, and Ferrin}{Pettersen et~al\mbox{.}}{2004}]%
        {paper29}
\bibfield{author}{\bibinfo{person}{Eric~F. Pettersen},
  \bibinfo{person}{Thomas~D. Goddard}, \bibinfo{person}{Conrad~C. Huang},
  \bibinfo{person}{Gregory~S. Couch}, \bibinfo{person}{Daniel~M. Greenblatt},
  \bibinfo{person}{Elaine~C. Meng}, {and} \bibinfo{person}{Thomas~E. Ferrin}.}
  \bibinfo{year}{2004}\natexlab{}.
\newblock \showarticletitle{{UCSF} Chimera - {A} visualization system for
  exploratory research and analysis}.
\newblock \bibinfo{journal}{{\em Journal of Computational Chemistry\/}}
  \bibinfo{volume}{25}, \bibinfo{number}{13} (\bibinfo{year}{2004}),
  \bibinfo{pages}{1605--1612}.
\newblock
\showDOI{%
\url{https://doi.org/10.1002/jcc.20084}}


\bibitem[\protect\citeauthoryear{Phillips}{Phillips}{2012}]%
        {paper31}
\bibfield{author}{\bibinfo{person}{Jeff~M. Phillips}.}
  \bibinfo{year}{2012}\natexlab{}.
\newblock \showarticletitle{Chernoff-Hoeffding Inequality and Applications}.
\newblock \bibinfo{journal}{{\em CoRR\/}}  \bibinfo{volume}{abs/1209.6396}
  (\bibinfo{year}{2012}).
\newblock
\showURL{%
\url{http://arxiv.org/abs/1209.6396}}


\bibitem[\protect\citeauthoryear{USDA}{USDA}{2017}]%
        {paper9}
\bibfield{author}{\bibinfo{person}{USDA}.} \bibinfo{year}{2017}\natexlab{}.
\newblock \bibinfo{title}{Annual Wheat production data,USA}.
\newblock   (\bibinfo{year}{2017}).
\newblock
\showURL{%
\url{https://www.ers.usda.gov/data-products/wheat-data/}}


\bibitem[\protect\citeauthoryear{Wood}{Wood}{2017}]%
        {paper21}
\bibfield{author}{\bibinfo{person}{Gavin Wood}.}
  \bibinfo{year}{2017}\natexlab{}.
\newblock \bibinfo{title}{ETHEREUM: A secure decentralized generalized
  transaction ledger}.
\newblock   (\bibinfo{year}{2017}).
\newblock
\showURL{%
\url{http://gavwood.com/paper.pdf}}


\end{thebibliography}


@article{paper1,
author="Jason, Paul, Cruz and Yuichi, Kaji",
title="The Bitcoin Network as Platform for Trans-Organizational Attribute Authentication",
journal="IPSJ SIG Notes",
ISSN="09196072",
publisher="Information Processing Society of Japan (IPSJ)",
year="2015",
month="feb",
volume="2015",
number="12",
pages="1-6",
URL="http://ci.nii.ac.jp/naid/110009877764/en/",
DOI="",
}
@article{paper2,
  title={MedRec: Medical Data Management on the Blockchain},
  author={Ariel Ekblaw, Asaf Azaria, Thiago Vieira, Andrew Lippman},
  year={2016},
  note={version: 57e013615dbf3f3300152554},
  publisher={PubPub},
}
@ONLINE{paper3,
author = {David Bollier},
title = {Reinventing Law for the Commons},
month = September,
year = {2015},
url = {http://www.commonaccord.org/}
}

@InProceedings{paper4,
  Title                    = {{Securing Video Integrity Using Decentralized Trusted Timestamping on the Blockchain}},
  Author                   = {{G}ipp, {B}ela and {K}osti, {J}agrut and {B}reitinger, {C}orinna},
  Booktitle                = {{P}roceedings of the 10th {M}editerranean {C}onference on {I}nformation {S}ystems ({MCIS})},
  Year                     = {2016},
  
  Address                  = {Paphos, Cyprus},
  Month                    = sep # { 4-6}
} 
@article{paper5,
  title={The importance of data set provenance for science},
  author={Hills, D. J., R. R. Downs, R. Duerr, J. C. Goldstein, M. A. Parsons, and H. K. Ramapriyan},
  year={2015},
  note={version: doi:10.1029/2015EO040557},
  publisher={Eos},
}
@Manual{paper6,
  title = 	 {Open Provenance Model},
  organization = {Open Provenance model},
  month =	 {August},
  year =	 2007,
  note =	 {\url{http://openprovenance.org/}}
}
@article{paper7,
  title={Better with bitcoin},
  author={The Economist},
  year={2016},
  note={http://www.economist.com/news/science-and-technology/21699099-blockchain-technology-could-improve-reliability-medical-trials-better},
  publisher={The Economist},
}
@article{paper8,
  title={A Next-Generation Smart Contract and Decentralized Application Platform},
  author={Vitalik Buterin},
  year={2015},
  note={September},
  publisher={PubPub},
}
@ONLINE{paper9,
author = {USDA},
title = {Annual Wheat production data,USA },
month = march,
year = {2017},
url = {https://www.ers.usda.gov/data-products/wheat-data/}
}
@MANUAL{paper10,
  title = 	 {NAMECOIN},
  author =	 {Vincent Durham},
  month =	 {September},
  year =	 2010,
  note =	 {\url{https://namecoin.org/}}
}
@misc{paper11,
    author = {Peter Linder},
    title = {DEcryption Contract ENforcement Tool (DECENT): A Practical Alternative to Government Decryption Backdoors},
    howpublished = {Cryptology ePrint Archive, Report 2016/245},
    year = {2016},
    note = {\url{http://eprint.iacr.org/2016/245}},}
@misc{paper12,
    author = {Roman Jagomägis and Peeter Laud and Alisa Pankova},
    title = {Preprocessing-Based Verification of Multiparty Protocols with Honest Majority},
    howpublished = {Cryptology ePrint Archive, Report 2015/674},
    year = {2015},
    note = {\url{http://eprint.iacr.org/2015/674}},
}
@misc{paper13,
    author = {Kevin Delmolino, Mitchell Arnett, Ahmed Kosba, Andrew Miller, and Elaine Shi},
    title = {Step by Step Towards Creating a Safe Smart Contract: Lessons and Insights from a Cryptocurrency Lab},
    howpublished = {Cryptology ePrint Archive, Report 2015/460},
    year = {2015},
    note = {\url{http://eprint.iacr.org/2015/460}},
}
@inproceedings{paper14,
  author = {Juels, Ari and Kosba, Ahmed E. and Shi, Elaine},
  booktitle = {ACM Conference on Computer and Communications Security},
  crossref = {conf/ccs/2016},
  editor = {Weippl, Edgar R. and Katzenbeisser, Stefan and Kruegel, Christopher and Myers, Andrew C. and Halevi, Shai},
  ee = {http://doi.acm.org/10.1145/2976749.2978362},
  interhash = {0c408d53f95548f79b3e57d25815e5f5},
  intrahash = {168efdbf79bf1d0a92a836c8dcd5e4db},
  isbn = {978-1-4503-4139-4},
  pages = {283-295},
  publisher = {ACM},
  title = {The Ring of Gyges: Investigating the Future of Criminal Smart Contracts.},
  url = {http://dblp.uni-trier.de/db/conf/ccs/ccs2016.html#JuelsKS16},
  year = 2016
}
@MANUAL{paper15,
  title = 	 {Clinical Research},
  author =	 {US Food and Drug Administration},
  month =	 {April},
  year =	 2017,
  note =	 {\url{https://www.fda.gov/ForPatients/Approvals/Drugs/ucm405622.htm}}
}
@inproceedings{paper16,
  author    = {Lei Xu and
               Lin Chen and
               Nolan Shah and
               Zhimin Gao and
               Yang Lu and
               Weidong Shi},
  title     = {{DL-BAC:} Distributed Ledger Based Access Control for Web Applications},
  booktitle = {Proceedings of the 26th International Conference on World Wide Web
               Companion, Perth, Australia, April 3-7, 2017},
  pages     = {1445--1450},
  year      = {2017},
  crossref  = {DBLP:conf/www/2017c},
  url       = {http://doi.acm.org/10.1145/3041021.3053897},
  doi       = {10.1145/3041021.3053897},
  timestamp = {Tue, 18 Apr 2017 15:34:36 +0200},
  biburl    = {http://dblp.uni-trier.de/rec/bib/conf/www/XuCSGLS17},
  bibsource = {dblp computer science bibliography, http://dblp.org}
  }
@article{paper17,
  author    = {Aafaf Ouaddah and
               Anas Abou El Kalam and
               Abdellah Ait Ouahman},
  title     = {FairAccess: a new Blockchain-based access control framework for the
               Internet of Things},
  journal   = {Security and Communication Networks},
  volume    = {9},
  number    = {18},
  pages     = {5943--5964},
  year      = {2016},
  url       = {http://dx.doi.org/10.1002/sec.1748},
  doi       = {10.1002/sec.1748},
  timestamp = {Fri, 21 Apr 2017 13:11:06 +0200},
  biburl    = {http://dblp.uni-trier.de/rec/bib/journals/scn/OuaddahKO16},
  bibsource = {dblp computer science bibliography, http://dblp.org}
}
@ONLINE{paper18,
author = {Project Provenance Ltd},
title = {Blockchain: the solution for transparency in product supply chains},
month = November,
year = {2015},
url = {https://www.provenance.org/whitepaper}
}
@ONLINE{paper19,
author = {Gideon Greenspan},
title = {Four Genuine Blockchain Use Cases},
month = May,
year = {2016},
url = {http://www.coindesk.com/four-genuine-blockchain-use-cases/}
}
@inproceedings{paper20,
  author    = {Matt Lepinski and
               Silvio Micali and
               Abhi Shelat},
  title     = {Fair-Zero Knowledge},
  booktitle = {Theory of Cryptography, Second Theory of Cryptography Conference,
               {TCC} 2005, Cambridge, MA, USA, February 10-12, 2005, Proceedings},
  pages     = {245--263},
  year      = {2005},
  crossref  = {DBLP:conf/tcc/2005},
  url       = {http://dx.doi.org/10.1007/978-3-540-30576-7_14},
  doi       = {10.1007/978-3-540-30576-7_14},
  timestamp = {Mon, 08 Aug 2011 15:22:27 +0200},
  biburl    = {http://dblp.uni-trier.de/rec/bib/conf/tcc/LepinskiMS05},
  bibsource = {dblp computer science bibliography, http://dblp.org}
}
@ONLINE{paper21,
author = {Gavin Wood},
title = {ETHEREUM: A secure decentralized generalized transaction ledger},
month = February,
year = {2017},
url = {http://gavwood.com/paper.pdf}
}
@ONLINE{paper22,
author = {MeteorJs},
title = Meteorjs,
month = May,
year = {2016},
url = {https://www.meteor.com/}
}
@article{paper23,
  author    = {George SL, Buyse M},
  title     = {Data fraud in clinical trials},
  journal   = {PMC},
  volume    = {5},
  number    = {2},
  pages     = {161-173},
  year      = {2015},
  url       = {http://dx.doi.org/10.1002/sec.1748},
  doi       = {10.4155/cli}
}
@MANUAL{paper24,
  title = 	 {Google Appscript},
  author =	 {Google },
  month =	 {April},
  year =	 2017,
  note =	 {\url{https://developers.google.com/apps-script/}}
}
@ONLINE{paper25,
author = {MongoDB},
title = {MongoDB},
month = Jan,
year = {2017},
url = {https://www.mongodb.com/}
}
@ONLINE{paper26,
author = {Jonathan Brown},
title = {Storing compressed text in Ethereum transaction logs},
month = October,
year = {2015},
url = {http://jonathanpatrick.me/blog/ethereum-compressed-text}
}
@misc{paper27,
    author = {Ahmed Kosba and Andrew Miller and Elaine Shi and Zikai Wen and Charalampos Papamanthou},
    title = {Hawk: The Blockchain Model of Cryptography and Privacy-Preserving Smart Contracts},
    howpublished = {Cryptology ePrint Archive, Report 2015/675},
    year = {2015},
    note = {\url{http://eprint.iacr.org/2015/675}},
}
@inproceedings{paper28,
  author    = {Zvika Brakerski and
               Jonathan Katz and
               Gil Segev and
               Arkady Yerukhimovich},
  title     = {Limits on the Power of Zero-Knowledge Proofs in Cryptographic Constructions},
  booktitle = {Theory of Cryptography - 8th Theory of Cryptography Conference, {TCC}
               2011, Providence, RI, USA, March 28-30, 2011. Proceedings},
  pages     = {559--578},
  year      = {2011},
  crossref  = {DBLP:conf/tcc/2011},
  url       = {http://dx.doi.org/10.1007/978-3-642-19571-6_34},
  doi       = {10.1007/978-3-642-19571-6_34},
  timestamp = {Wed, 30 Mar 2011 13:43:54 +0200},
  biburl    = {http://dblp.uni-trier.de/rec/bib/conf/tcc/BrakerskiKSY11},
  bibsource = {dblp computer science bibliography, http://dblp.org}
}
@article{paper29,
  author    = {Eric F. Pettersen and
               Thomas D. Goddard and
               Conrad C. Huang and
               Gregory S. Couch and
               Daniel M. Greenblatt and
               Elaine C. Meng and
               Thomas E. Ferrin},
  title     = {{UCSF} Chimera - {A} visualization system for exploratory research
               and analysis},
  journal   = {Journal of Computational Chemistry},
  volume    = {25},
  number    = {13},
  pages     = {1605--1612},
  year      = {2004},
  url       = {http://dx.doi.org/10.1002/jcc.20084},
  doi       = {10.1002/jcc.20084},
  timestamp = {Tue, 07 Sep 2004 14:08:20 +0200},
  biburl    = {http://dblp.uni-trier.de/rec/bib/journals/jcc/PettersenGHCGMF04},
  bibsource = {dblp computer science bibliography, http://dblp.org}
}
@article{paper30,
  author    = {Tim Clark and
               Paolo Ciccarese and
               Carole A. Goble},
  title     = {Micropublications: a Semantic Model for Claims, Evidence, Arguments
               and Annotations in Biomedical Communications},
  journal   = {CoRR},
  volume    = {abs/1305.3506},
  year      = {2013},
  url       = {http://arxiv.org/abs/1305.3506},
  timestamp = {Sun, 02 Jun 2013 20:48:21 +0200},
  biburl    = {http://dblp.uni-trier.de/rec/bib/journals/corr/abs-1305-3506},
  bibsource = {dblp computer science bibliography, http://dblp.org}
}
@article{paper31,
  author    = {Jeff M. Phillips},
  title     = {Chernoff-Hoeffding Inequality and Applications},
  journal   = {CoRR},
  volume    = {abs/1209.6396},
  year      = {2012},
  url       = {http://arxiv.org/abs/1209.6396},
  timestamp = {Wed, 10 Oct 2012 21:28:55 +0200},
  biburl    = {http://dblp.uni-trier.de/rec/bib/journals/corr/abs-1209-6396},
  bibsource = {dblp computer science bibliography, http://dblp.org}
}
@book{paper32,
 author = {Goldreich, Oded},
 title = {Foundations of Cryptography: Volume 2, Basic Applications},
 year = {2004},
 isbn = {0521830842},
 publisher = {Cambridge University Press},
 address = {New York, NY, USA},
}
@inproceedings{paper33,
 author = {Liang, Xueping and Shetty, Sachin and Tosh, Deepak and Kamhoua, Charles and Kwiat, Kevin and Njilla, Laurent},
 title = {ProvChain: A Blockchain-based Data Provenance Architecture in Cloud Environment with Enhanced Privacy and Availability},
 booktitle = {Proceedings of the 17th IEEE/ACM International Symposium on Cluster, Cloud and Grid Computing},
 series = {CCGrid '17},
 year = {2017},
 isbn = {978-1-5090-6610-0},
 location = {Madrid, Spain},
 pages = {468--477},
 numpages = {10},
 url = {https://doi.org/10.1109/CCGRID.2017.8},
 doi = {10.1109/CCGRID.2017.8},
 acmid = {3101176},
 publisher = {IEEE Press},
 address = {Piscataway, NJ, USA},
 keywords = {Blockchain, Blockchain Cloud, Cloud Computing, Data provenance, Privacy, Reliability},
}

\end{document}